\documentclass[onecollarge]{svjour2}       
\smartqed  
\usepackage{amsmath}
\usepackage[dvipdfmx]{graphicx}

\usepackage{color}
\usepackage{multirow}
\makeatletter
\AtBeginDocument{\@ifpackageloaded{natbib}{\ifNAT@numbers\if@filesw\immediate\write\@auxout{\string\global\string\NAT@numberstrue}\fi\fi}{}}
\makeatother
\begin{document}

\title{Generalised central limit theorems for growth rate distribution of complex systems}


\author{Misako Takayasu \and Hayafumi Watanabe \and Hideki Takayasu}


\institute{Misako Takayasu \and Hayafumi Watanabe$^{*}$  
           \at Department of Computational Intelligence \& Systems Science, Interdisciplinary Graduate School of Science \& Engineering, Tokyo Institute of Technology, 4259-G3-52 Nagatsuta-cho, Midori-ku, Yokohama 226-8502, Japan \\
           \and
           Hideki Takayasu
           \at
           Sony Computer Science Laboratories Inc., 3-14-13 Higashigotanda, Shinagawa-ku, Tokyo 141-0022, Japan \\
           Meiji Institute for Advanced Study of Mathematical Sciences, 1-1-1 Higashimita, Tama-ku, Kawasaki 214-8571, Japan \\
            \email{h-watanabe@smp.dis.titech.ac.jp}$^{*}$
           }
         
             %
             %
            %


\date{Received: date / Accepted: date}

\maketitle

\begin{abstract}
We introduce a solvable model of randomly growing systems consisting of many independent subunits. Scaling relations and growth rate distributions in the limit of infinite subunits are analysed theoretically. Various types of scaling properties and distributions reported for growth rates of complex systems in a variety of fields can be derived from this basic physical model. Statistical data of growth rates for about 1 million business firms are analysed as a real-world example of randomly growing systems. Not only are the scaling relations  consistent with the theoretical solution, but the entire functional form of the growth rate distribution 
is fitted with a theoretical distribution that has a power-law tail.
\textcolor{black}{}
\end{abstract}

\section{Introduction}
\label{Eq:intro}
 In general, growth phenomena are highly irreversible dynamical processes far from thermal equilibrium \cite{R:1}.
\begin{figure}[t]
\centering
\begin{minipage}{0.8\hsize}
 \includegraphics[width=13.5cm]{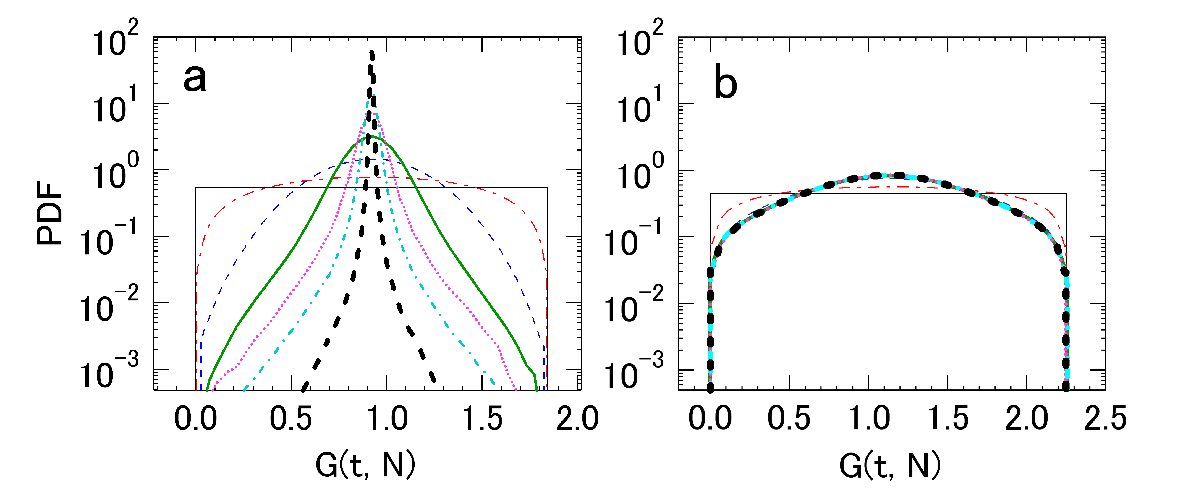}
\end{minipage}
\centering
\begin{minipage}{0.4\hsize}
  \includegraphics[width=7.3cm]{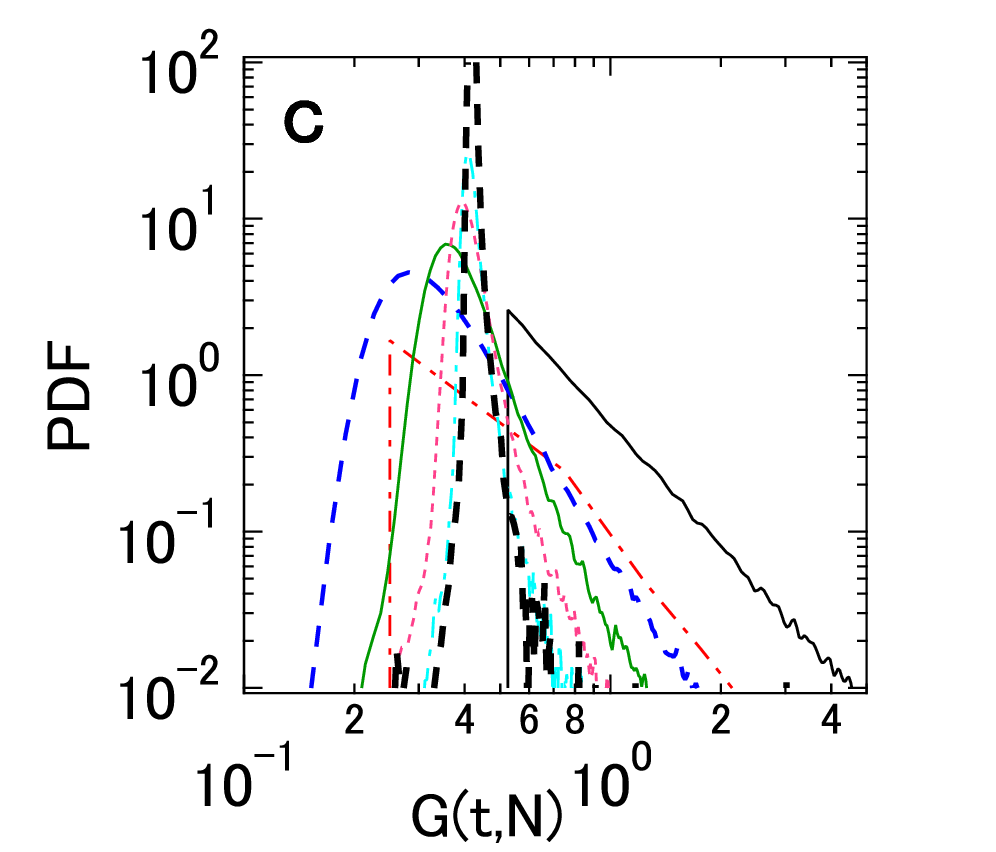}
\end{minipage}
\centering
\begin{minipage}{0.4\hsize}
  \includegraphics[width=7.3cm]{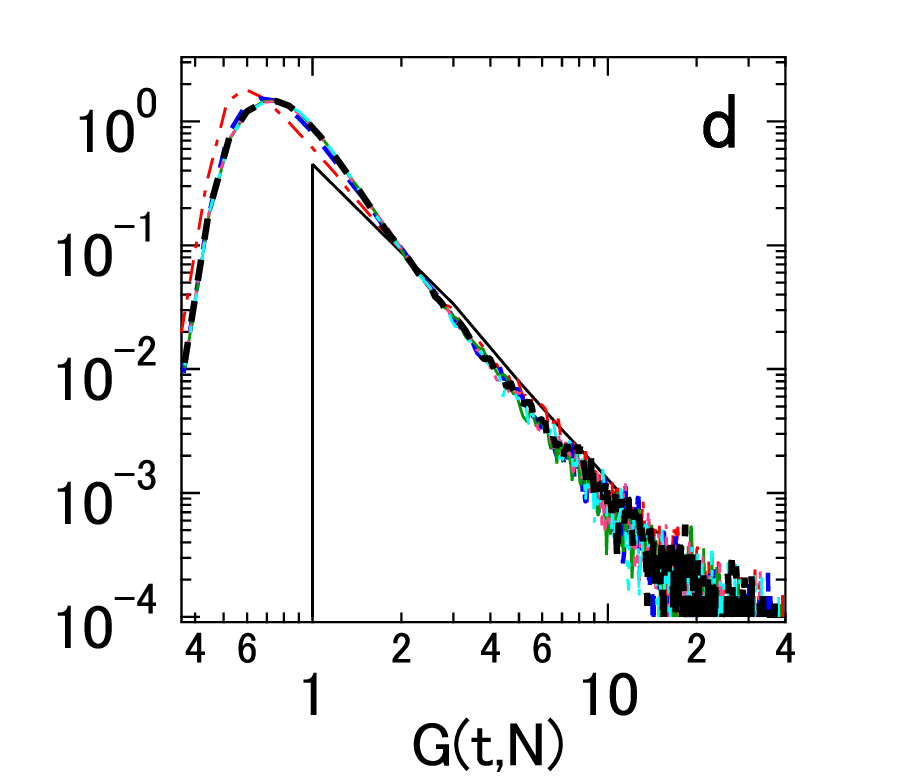}
\end{minipage}
\caption{Semi-logarithmic plots, comparing system size dependence of the probability density functions of growth rates for  (a)  $\alpha=1.5$ ($\langle g_j(t)^{1.5}\rangle =1$) and (b) $\alpha=0.5$ ($\langle g_j(t)^{0.5}\rangle =1$). In both cases, the growth rate  of individual subunits follows a uniform distribution. Plotted are  systems with $N=1$ (black thin line), $N=2$ (red dash-dotted line), $N=10$ (blue broken line), $N=10^2$ (bold green line), $N=10^3$ (purple dotted line), $N=10^4$ (light blue dash-dotted line), and  $N=10^6$ (black broken line). \textcolor{black}{Figures corresponding to the power-law distribution $P(>g_i(t)) \propto g_i^{-1.6}$ are shown for  $\langle g_i(t)^{1.5}\rangle=1$ in panel (c) and for $\langle g_i(t)^{0.5}\rangle =1$ in  panel (d).}  
}
\label{fig:f1}       
\end{figure}
%
\begin{figure}[t]
\begin{minipage}{0.5\hsize}
 \includegraphics[width=1\textwidth]{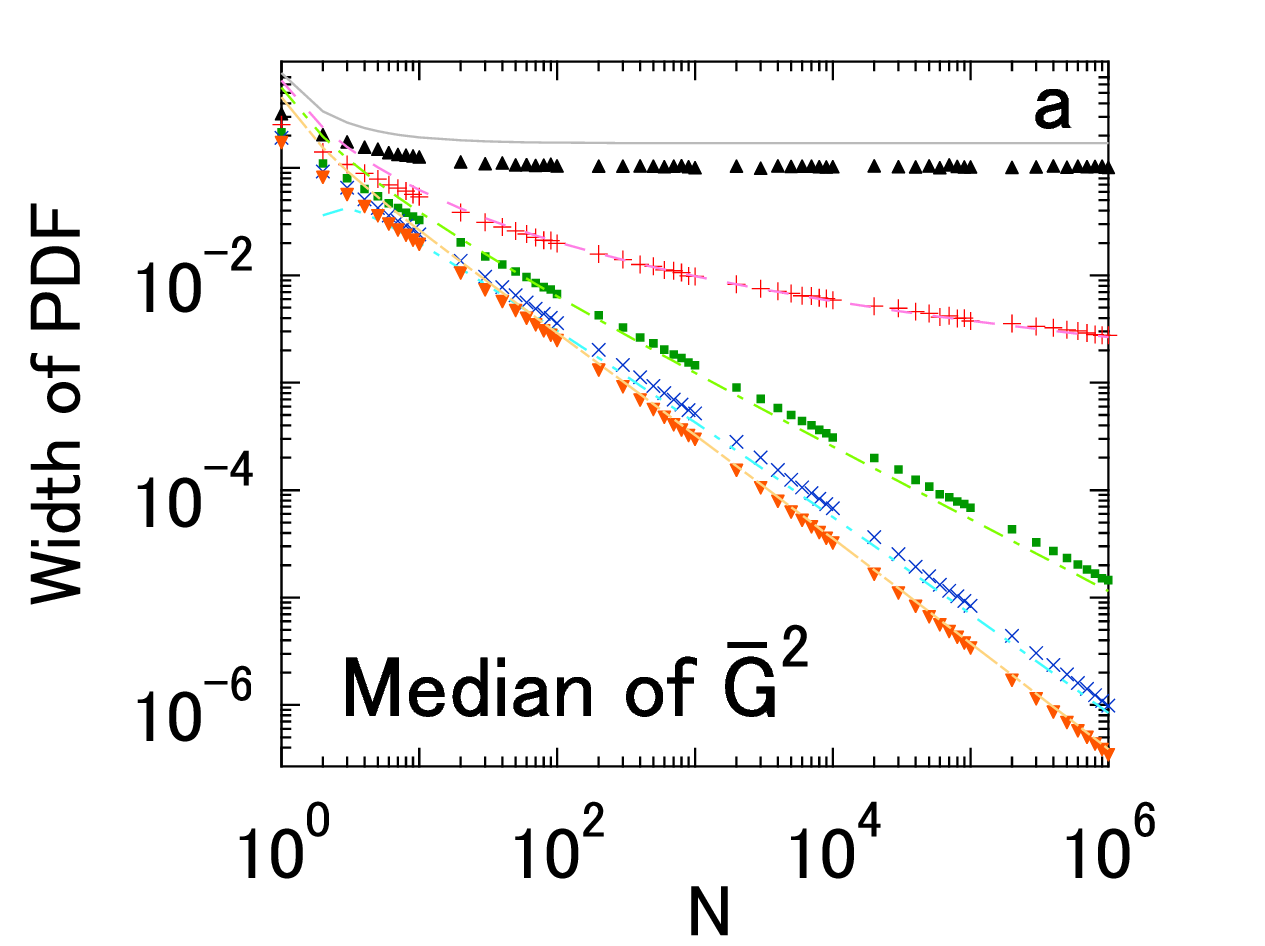}
\end{minipage}
\begin{minipage}{0.5\hsize}
   \includegraphics[width=1\textwidth]{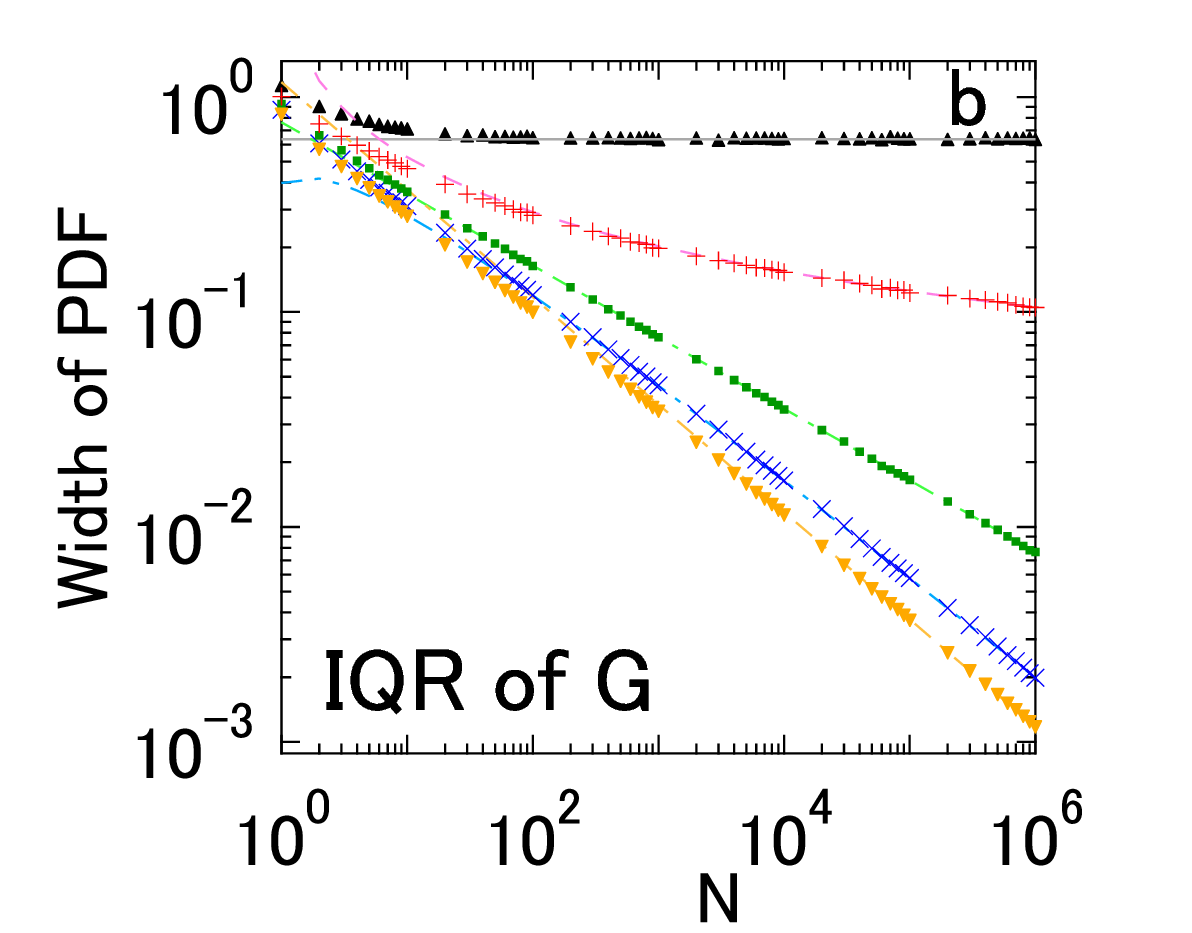}
\end{minipage}
\caption{
Confirmation of theoretical estimates of size dependence of the width of the growth rate distribution.
The widths of growth rate distributions are measured by repeating numerical simulations for various values of $N$. The points are plotted on a log--log scale for five cases: $\alpha =0.5$, $1.0$, $1.5$, $2.0,$ and $2.5$ (black, red, green, blue, and orange, respectively), with simulation results  marked as points and theory  as lines. \textcolor{black}{For numerical simulations, the growth rate for each subunit, $g_j(t)$, is distributed uniformly in the range $(0,(\alpha+1)^{1/\alpha}]$ satisfying $\langle g_j(t)^{\alpha}\rangle =1$ ($\alpha=0.5$, $1.0$, $1.5$, $2.0,$ and $2.5$).
(a) Data for the median of $\bar{G}^2$ for numerical simulations and Eq. (\ref{Eq:c8}) for theoretical estimations.
(b) Data for the interquartile range (IQR) of $G$ for numerical simulations and the square roots of the asymptotic functional forms of Eq. (\ref{Eq:c8}) given in Table 1 for theoretical estimations. The theoretical curves are shifted  along the vertical axis so that they overlap with the numerical simulations at $N=10^6$ in  panel (b).  The IQR, which is  one of the most commonly used robust measures of width of a probability density function (PDF), is the difference between the largest and smallest values in the middle 50\% of a set of data. 
From these figures, we can confirm that the medians of $\bar{G}(t;N)^2$ almost correspond to the width given by Eq. (\ref{Eq:c8}), and IQRs are proportional to the asymptotic behaviour of this equation given by Table 1.}
}
\label{fig:f2}       
\end{figure} 
 From the viewpoint of statistical physics, it is an important new topic that growth rates of complex systems often show nontrivial similar statistical behaviours across fields of sciences. Fat-tailed distributions of growth rates and a nontrivial decrease of variance as a function of  size are reported in various fields of sciences and are seen in data for business firms \cite{R:3,R:4,R:5,R:2}, sales of pharmaceutical products \cite{R:6}, circulation numbers of newspapers \textcolor{black}{\cite{R:7}}, population of migratory birds \cite{R:8}, animal metabolic rate fluctuations \cite{R:9}, the amount of scientific funding \cite{R:10}, group size of religious activities \cite{R:11}, population size of cities \cite{R:12},  national economic activity  GDP \cite{R:13}, and the amount of exports and government debt \cite{R:14}. The probability densities of \textcolor{black}{logarithmic} growth rates in most of these examples are typically approximated by double-exponential (Laplace) distributions or by power-law distributions---quite interestingly, not by a Gaussian distribution.  \par
     Statistics on growth rates of business firms have a long history of study, and recently statistical physicists have become involved in this topic. Gibrat postulated  the ``law of proportional effect'': The expected value of the growth rate of a business firm is proportional to the current size of the firm \cite{R:15,R:14}. In Gibrat's original assumption, he states that the variance of growth rates is independent of the size; however, data analyses of business firm activities show various types of variance--size relations. There are papers that support  Gibrat's original  assumption \cite{R:17,R:16}; however, nontrivial fractional power laws are reported not only for business firms but also for many other phenomena \cite{R:3,R:4,R:5,R:6,R:12,R:8,R:9,R:13,R:11,R:7,R:10,R:2,R:1}. Further, dependence on country \cite{R:18} and the transition from Gibrat's assumption to such power-law decays have been pointed out \cite{R:19}. Various types of theoretical models of business firms have been introduced by physicists for better understanding of their scaling properties  from the standpoint of complex systems \cite{R:21,R:24,R:26,R:23,R:42,R:20,R:25,R:39,R:43,R:41,R:22}; however, there has been no unified theory that can explain all these basic properties simultaneously. \textcolor{black}{To understand these phenomena consistently, we introduce a simple random growth model of a complex system consisting of many independent subunits, and we consider the relationship between fluctuations in the growth rate of subunits with that of total system.} \par
 \textcolor{black}{There are a number of pioneering studies on complex systems that focus on  growth rate statistics. Wyart et al. introduced a company model that consisted of independent subunits characterised by a power-law size distribution \cite{wyart2003statistical}.  The growth rate of the overall system was shown to observe a symmetric stable distribution with its scale parameter, which corresponds to the standard deviation, either decaying in accordance with a power law or converging to a constant in the limit of infinite unit numbers. Schwarzkopf et al. investigated a model that consisted of independent subunits whose number of summations changes with time, and they showed that the growth rate of this model also observed a stable distribution  \cite{schwarzkopf2010cause}.
Malevergne et al. introduced a theoretical model of firms based on the birth and death process and studied the contribution of the growth rate of the entire economy to the power-law exponent of the distribution of sales \textcolor{black}{\cite{malevergne2008zipf,malevergne2013zipf}}.
 Solomon et al. studied the statistical properties of growth rates in the framework of the nonlinear dynamics of a generalised Lotka--Volterra model  \cite{biham1998generic,solomon1998stochastic,huang2002stochastic,solomon2002stable}. They demonstrated that the growth rate distribution  for large-time-scale windows observes a stable distribution. } \par
  In the next section, \textcolor{black}{we introduce a basic model of a complex system whose growth rate can be theoretically derived by a kind of renormalisation of the many independent subunits of which the system consists.} In  Section 3, we show that in the limit of an infinite number of subunits, the distribution of growth rates is shown to converge to a stable distribution with a power-law tail \textcolor{black}{and its scale indicator shrinks in a nontrivial manner.} The stable distribution and the generalised central limit theorem were established in mathematics about 80 years ago \cite{R:28,R:27}; however, these concepts were applied mostly to theoretical models by assuming scaling properties for phenomena such as turbulence \cite{R:29,R:30}. The validity of the theoretical results for growth rates is confirmed in Section 4 by analysing a huge database on business firms. This example may be the first real-world application of an asymmetric stable distribution fitted for the whole scale range. The final section is devoted to discussion. \par
\section{Model}
\label{Eq:sec:1}
 We consider a system consisting of $N$ subunits characterised by non-negative scalar quantities, $\{x_j(t)\}$. For each subunit, we assume the following random multiplicative time evolution \cite{R:31}, which is known to be one of the basic processes for producing power-law fluctuations \cite{R:33,R:32}:  
\begin{equation}
x_j(t+\Delta t)=g_j(t)x_j(t)+f_j(t), \label{Eq:c1}
\end{equation}
where $g_j(t)$ and $f_j(t)$ for $j=1,2,\dots,N$ are growth rates and external forces, respectively, and both are assumed to be independent identically distributed random variables taking only positive values. (See Appendix A for a brief review of random multiplicative processes).  In the case in which the probability of occurrence of $g_j(t) > 1 $ is not zero and if $\langle \log (g_j(t)) \rangle  < 0$, where $\langle \cdots\rangle $ denotes the average, it is known that there exists a statistically steady state in which the cumulative distribution follows a power law \cite{R:31},
\begin{equation}
P(>x_j) \propto x_j^{-\alpha} ,  \label{Eq:c2}      
\end{equation}
for large value of $x_j$ with the positive exponent   determined exactly only by the statistics of the growth rate by solving the following equation: 
\begin{equation}
\langle g_j(t)^\alpha\rangle =1.  \label{Eq:c3}
\end{equation}                                 
\textcolor{black}{We note that Solomon et al. elaborated upon a generalised version of this type of  model, $x(t+1)=g(t)f_1(x(t))+f_2(x(t)) ,$ where both $f_1$ and $f_2$ are nonlinear functions, in which a power-law distribution holds generally in the steady state, although the simple exact relation given in Eq. (\ref{Eq:c3}) no longer holds \cite{solomon2002stable}. } \par
  From a renormalisation point of view we pay attention to the sum of all subunits, $X(t; N) \equiv \sum^{N}_{j=1}x_j(t)$, 
  which follows the same type of time evolution as that of the subunits:
\begin{equation}
X(t+\Delta t; N)=G(t; N)X(t; N)+F(t; N), \label{Eq:c4}
\end{equation}
where $F(t; N) \equiv \sum^{N}_{j=1}f_j(t)$, and the growth rate of the whole system is defined as
\begin{equation}                                            
G(t;N) \equiv \frac{\sum^N_{j=1}g_j(t)x_j(t)}{\sum^{N}_{j=1}x_j(t)}. \label{Eq:c5}
\end{equation}
It is easy to show that the mean value of growth rates is invariant, namely, $\langle G(t;N)\rangle =\langle g_j(t)\rangle  \equiv G$. \par
\textcolor{black}{The system growth rate, $G(t;N)$, in this study is closely related to the company growth model introduced elsewhere in which the probability density of the size change of a firm, under the assumption that a firm is a composite of $K$  independent subunits, 
 is given by the following \cite{wyart2003statistical}:
\begin{equation}
Q(S,R)=\sum^{\infty}_{K=1}L(K) \int ds_i \prod^{K}_{i=1}P_{s}(s_i) \delta(S-\sum_{i=1}^{K}s_i) 
\int d\eta_i \prod_{i=1}^{K}P_{\eta}(\eta_i) \delta(R-\sum_{i=1}^{K}s_i \eta_i),  
\end{equation}
where $S$ is the firm size for the previous term, $R$ is that for the current term, 
$P_{s}(s)$ is the  probability density function (PDF) of the sizes of subunits supposing $P_{s}(s) \propto s^{-\alpha-1}$, $P_{\eta}(\eta)$ is the PDF of the growth rate of the subunits, $L(K)$ is the number density of subunits, and $\delta(x)$ is the Dirac delta function for continuous variables and the Kronecker delta for discrete variables. 
Our model's growth rate $G(N, t)$ in the steady state corresponds to the ratio $R/S$  for the case of fixed subunit number,  $L(K)=\delta(K-N)$. 
In contrast, the authors in Ref. \cite{wyart2003statistical} mainly discussed the growth rate of the model on the condition $L(K)=1$.
} 
\textcolor{black}{In terms of this class of system growth models, our research newly clarifies the dependence of the distribution of the system growth rate \textcolor{black}{on both the number of subunits and  the distribution of subunit growth rates.}
In our study, we 
assume that $\{g_i\}$ are identically distributed independent variables defined on a non-negative range, whereas in Ref. \cite{wyart2003statistical} $\{\eta_i\}$ are assumed to follow a normal distribution with zero mean to calculate the distribution of the system growth rate $G$.}
\par
 The properties of this model can be investigated by numerical simulation. Figure \ref{fig:f1}(a) shows an example of the deformation of growth rate distributions for various values of $N$ in the case in which Eq. (\ref{Eq:c3}) is fulfilled with $\alpha=1.5$ observed in the statistically steady state realised for time steps larger than $10^6$. Here the distribution of growth rates of subunits is given by an independent uniform distribution, as shown by the case of $N=1$. In this figure, the additive noise term, $f_j(t)$, is set to be a positive constant for simplicity, as we confirmed that the main results do not depend on the values of $f_j(t)$ except when it is identically $0$. As seen from this figure the distribution of growth rates of the aggregated system tends to shrink slowly to a delta function as $N$ goes to infinity. It is numerically confirmed that the same property of convergence to the delta function holds for any distribution of growth rates of subunits if the growth rate distribution satisfies Eq. (\ref{Eq:c3}) with $\alpha \geq 1$. \par
  Figure \ref{fig:f1}(b) shows an example for $\alpha=0.5$. In this case, the growth rate distribution stops shrinking for $N$ larger than $10$ and it converges to a nontrivial distribution in the limit of $N$ going to infinity. It is confirmed that this property is always observed if the value of $\alpha$ in Eq. (\ref{Eq:c3}) lies between $0$ and $1$. The distribution of the growth rate in the limit of $N = \infty $ depends on the functional form of the growth rate distribution for the subunits.  \par
   \textcolor{black}{Figures \ref{fig:f1}(c) and 1(d) show results indicative of the case where $g_i(t)$ observes a power-law distribution. 
 From these figures, we can also confirm that the distributions shrink for large $N$, in the case of $\alpha=1.5$. However, in contrast, the distribution stops shrinking for $N$ larger than $10$ in the case of $\alpha=0.5$. }
\section{ Theoretical analyses}
\label{Eq:sec:2}
\begin{figure}[t]
 \centering
 \begin{minipage}{0.8\hsize}
  \includegraphics[width=0.75\textwidth]{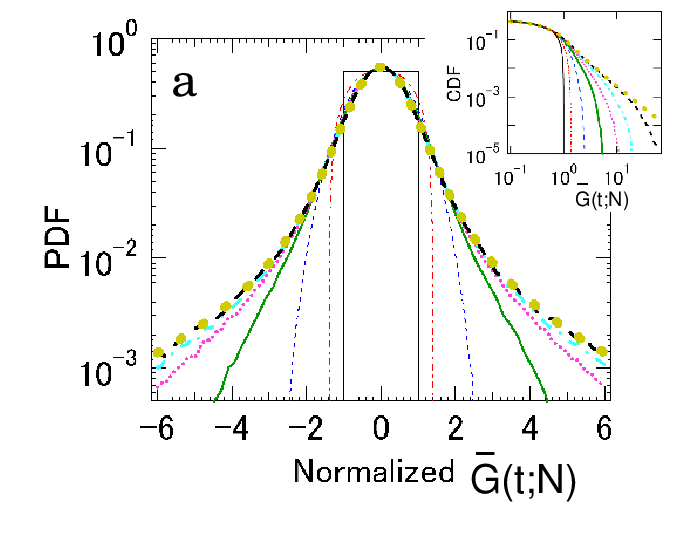}
 \end{minipage}
 \begin{minipage}{0.9\hsize}
  \includegraphics[width=0.75\textwidth]{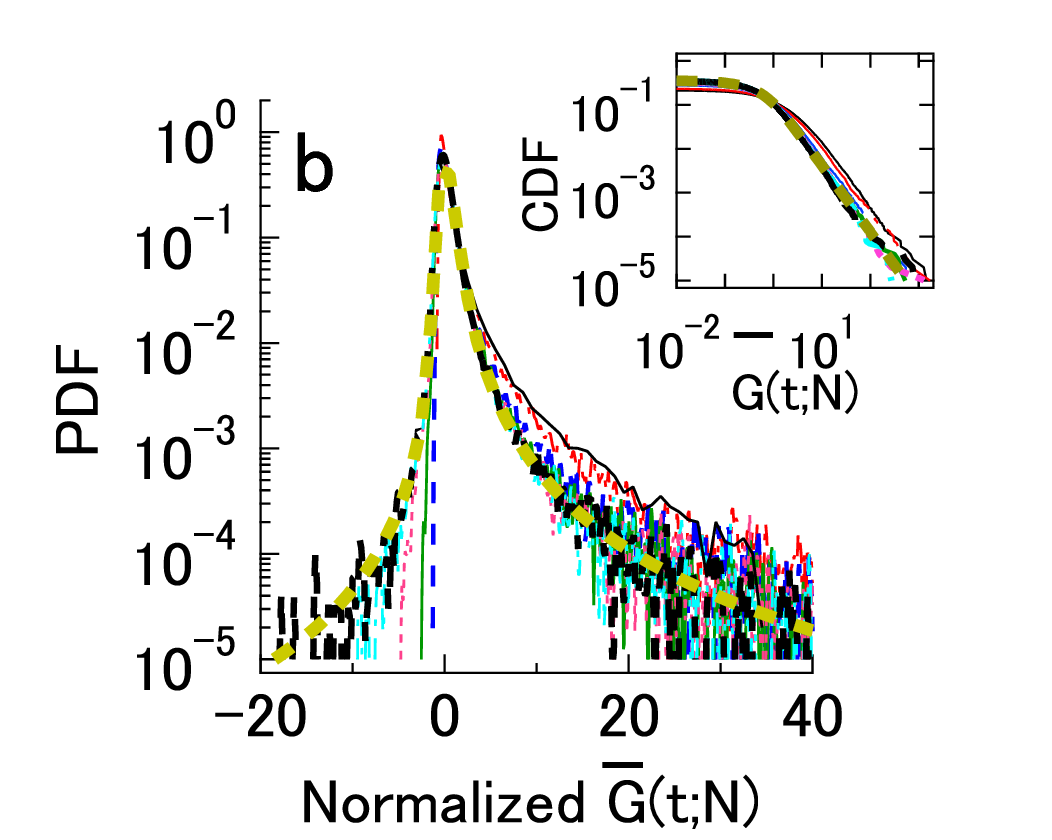}
 \end{minipage}
\caption{Convergence of the normalised growth rate distributions in the case of $\alpha=1.5$.  The growth rate for each subunit, $g_j(t)$, is distributed uniformly in the range $(0,(2.5)^{2/3}]$ satisfying $\langle g_j(t)^{1.5}\rangle =1$. The renormalised growth rate's PDFs for systems with $N$ subunits are plotted for several values of $N$: $N=1$ (black thin line), $N=2$ (red dash-dotted line), $N=10$ (blue broken line), $N=10^2$ (bold green line), $N=10^3$ (purple dotted line), $N=10^4$  (light blue dash-dotted line), and $N=10^6$ (black broken line);  the theoretical symmetric stable distribution $p(\bar{G};1.5,0)$  is plotted as an olive dotted line. The inset shows the cumulative distribution function of the positive part on a log--log scale, confirming convergence to the power law.
\textcolor{black}{Plots corresponding to the case of a power-law distribution of subunits, $P(>g_i(t)) \propto g_i^{-1.6}$, are shown for $\langle g_i(t)^{1.5}\rangle =1$ in panel (b) and the theoretical asymmetric stable distribution, $p(\bar{G};1.5,0.85)$, is also plotted by the olive dotted line for comparison.} 
}
\label{fig:f3}       
\end{figure}

\begin{table}
\begin{tabular}{|c||c|c|}
\hline
Value of $\alpha$ & \textcolor{black}{Width--Size($N$)} relation &\multirow{2}{*}{Limit distribution of growth rates}  \\
          $\langle g_j(t)^\alpha\rangle =1$        &(in the limit of $N \to \infty$ )& \\
\hline
\hline
$ 2 < \alpha$ &   \multirow{2}{*}{$ \frac{(\alpha-1)}{\sqrt{\alpha(\alpha-2)}}N^{-0.5} \to 0$} &  \multirow{4}{*}{Gaussian}  \\ 
$ \langle g_j(t)^2\rangle  < 1 $  &&\\ 
\cline{1-2}
$ \alpha=2$ &  \multirow{2}{*}{$(\frac{\gamma+\log (N)}{4})^{0.5} \cdot N^{-0.5} \to 0$} &   \\
$ \langle g_j(t)^2\rangle  = 1 $  && \\ 
\hline
$1< \alpha < 2$ &  \multirow{2}{*}{$\zeta{(\frac{2}{\alpha})^{0.5} \cdot (1-\frac{1}{\alpha})} \cdot N^{\frac{1}{\alpha}-1} \to 0$} &   \multirow{2}{*}{Stable distribution with power-law tails}   \\
$ \langle g_j(t)\rangle  < 1$, $\langle g_j(t)^2\rangle  >  1 $ && \\ 
\cline{1-2}
$ \alpha =1 $ &  \multirow{2}{*}{$\frac{\zeta{(2)}^{0.5}}{(\gamma+\log(N))} \to 0$} &   \multirow{2}{*}{$P(>|\bar{G}|\propto |\bar{G}|^{-\alpha})$}   \\
$ \langle g_j(t)\rangle  = 1 $  && \\ 
\cline{1-3}
\multirow{2}{*}{$0< \alpha <1$}& \multirow{2}{*}{$\frac{\zeta{(2/\alpha)}^{0.5}}{\zeta(1/\alpha)}$: constant } & \multirow{2}{*}{Nonuniversal distribution} \\
 && \\ 
$ \langle g_j(t)\rangle  > 1$, $\langle \log(g(t))\rangle  < 0$&(Gibrat's assumption holds) & depending on the subunit's properties \\  
\hline
\end{tabular}
\caption{Summary of generalised central limit theorems for growth rates, listing the value of $\alpha$ and 
\textcolor{black}{the asymptotic functional form of the square root of Eq. (\ref{Eq:c8}) divided by $\sigma$, which corresponds to the width of the growth rate, for large system size, $N$,} 
and the limit distributions. $\gamma=0.577\cdots$ is the Euler constant.}
\label{table:t1}
\end{table}

 Here, we can theoretically evaluate the $N$ dependence of \textcolor{black}{the width of the PDF of $G(t;n)$} \textcolor{black}{intuitively} by introducing an approximation of  random variables $\{x_j(t)\}$ that are known to follow a power law in the steady state. \textcolor{black}{A more precise derivation using the generalised central limit theorem is given in Appendix D.}
 \textcolor{black}{We introduce the measure of the width, $\bar{G}(t;N)^2$, where}
\textcolor{black}{
 we define $\bar{G}(t;N)$ as
\begin{equation}
\bar{G}(t;N) \equiv G(t;N)-G = \frac{\sum^{N}_{j=1}(g_j(t)-G) \cdot x_j(t)}{ \sum^{N}_{j=1}x_j(t) }. \label{Eq:c5b}
\end{equation}
$\bar{G}(t;N)^2$ takes zero in the case in which the PDF of $G(t;N)$ is a delta function.}
 Let $u$ be a random variable following a uniform distribution in the interval $(0, 1]$. Then, the distribution of the new variable, $u^{-1/\alpha}$,  follows a power law with exponent $\alpha$. For uniform random variables $\{u\}$, we can approximate $N$ samples by a set of deterministic values, $\left\{1/N, 2/N, \dots, 1 \right\}$, so that the set of power-law variables $\{x_j(t)\}$, which follow Eq. (\ref{Eq:c2}), can be approximated by the deterministic set  $\left\{N^{1/\alpha},(N/2)^{1/\alpha}, (N/3)^{1/\alpha}, \dots ,(N/j)^{1/\alpha}, \dots, 1 \right\}$. \textcolor{black}{Therefore, the typical sample of $\bar{G^2}(t;N)$ is obtained as follows: }
\begin{equation}
\bar{G_{r}}(t;N)^2 \equiv  \frac{\{\sum^{N}_{j=1}(g_j(t)-G) \cdot (j/N)^{-1/\alpha}\}^2}{\left\{\sum^{N}_{j=1}(j/N)^{-1/\alpha}\right\}^2}.
\end{equation}
\textcolor{black}{Taking the average of $\bar{G}_{r}(t;N)$ with respect to $g_j(t) (j=1,2,\dots ,N)$ and applying the independence condition, $\langle (g_n(t)-G)(g_m(t)-G)\rangle =\sigma^2\delta_{nm}$, we have
\begin{equation}
\langle \bar{G_{r}}(t;N)^2\rangle  = \sigma^2 \cdot \frac{\sum^{N}_{j=1}(j/N)^{-2/\alpha}}{\left\{\sum^{N}_{j=1}(j/N)^{-1/\alpha}\right\}^2}, \label{Eq:c6c}
\end{equation}
 where $\sigma^2$ is the variance of the growth rates for the subunits and $\delta_{nm}$ is Kroneker's delta.
}
 Then, we can calculate the summations in Eq. (\ref{Eq:c6c}), $N^{1/\alpha}\sum^{N}_{j=1}j^{-1/\alpha} $ and $N^{2/\alpha}\sum^{N}_{j=1}j^{-2/\alpha}$, by applying an asymptotic expansion formula for the Riemann Zeta function,

\begin{equation}
\zeta(\lambda) \equiv \sum^{\infty}_{j=1}\frac{1}{j^{\lambda}}=\sum^{N}_{j=1}\frac{1}{j^\lambda}+\frac{1}{(\lambda-1)N^{\lambda-1}}-\frac{1}{2N^{\lambda}}+\cdots . \label{Eq:c7}
\end{equation}
Neglecting the third term and higher order terms in the right-hand side of Eq. (\ref{Eq:c7}), we have the following approximation of Eq. (\ref{Eq:c6c}) for $0 < \alpha $: 
\begin{equation}
\left\langle \bar{G_r}(t,N)^2 \right> =\sigma^2 \cdot \frac{\zeta(\frac{2}{a})-\frac{\alpha}{2-\alpha}N^{1-\frac{2}{\alpha}}}{\left\{\zeta(\frac{1}{\alpha})-\frac{\alpha}{1-\alpha}N^{1-\frac{1}{\alpha}}\right\}^2} \label{Eq:c8}                    .             
\end{equation}
\par
 These theoretical evaluations are checked numerically in Fig. \ref{fig:f2}, in which the widths of growth rate distributions for the aggregated system are plotted as functions of the number of subunits, $N$. 
 %
 The theoretical asymptotic functional forms in Table 1 fit quite well asymptotically for all cases. 
\textcolor{black}{It should be noted that the standard deviation (or the variance) observes a slightly different scaling from that given Eq. (\ref{Eq:c8}) and is not suitable for the scale parameters of our model, except for extremely large $N$ (Appendix E).}
 \textcolor{black}{As an alternative to the standard deviation, we introduce the median of $\bar{G}(t;N)^2$ and the IQR as the measure of the width of the distribution in Figs. \ref{fig:f2}(a) and \ref{fig:f2}(b), respectively.}
\textcolor{black}{From Fig. \ref{fig:f2}(a) we can confirm that the medians of $\bar{G}(t;N)^2$ almost correspond to the width given by Eq. (\ref{Eq:c8}), and from Fig. \ref{fig:f2}(b) we see that the IQRs are proportional to the asymptotic behaviour of this equation given by Table 1. }\textcolor{black}{In addition, from Table 1 and Appendix D, we can also confirm that Eq. (\ref{Eq:c8}) is proportional to the scale parameter of the stable distribution given by Eq. (\ref{parameter3}) for $N\gg 1$.}
\par
From Eq. (\ref{Eq:c8}) we find that \textcolor{black}{the width} of the PDF of $G(t;N)$ converges to $0$ following a power law of $N$ in the case of $1 < \alpha$; in contrast,  \textcolor{black}{the width} takes a finite value even in the limit of $N \to \infty$ in the case of $0 < \alpha < 1 $. At the marginal case of $\alpha=1,$ the width decays to $0$ very slowly for increasing $N$. For $2< \alpha$ the \textcolor{black}{width} decays inversely proportional to $N$; that is, it obeys the typical $N$ dependence in the case of the ordinary central limit theorem. These functional forms of the $N$ dependence are summarised in Table 1 in the second column. From this result, we conclude that Gibrat's assumption of constant \textcolor{black}{variance} is fulfilled in the case of $0< a<1$ and that the nontrivial power-law decays of width--size relations observed in many complex systems are realised in the case of $1 < \alpha < 2$.   
     As shown in the first column of Table 1, the range of $\alpha$ is characterised by the form of equality or inequality for the moment function of the growth rates, $M(s) \equiv \langle g_j(t)^s\rangle $. The derivation of these relations is summarised in Appendix B along with the basic properties of the moment function. From this table we find that Gibrat's assumption holds for systems in which the mean growth rate is larger than $1$, whereas the nontrivial power-law decay of \textcolor{black}{the width--size} relation for the growth rate is expected in the situation in which the mean growth rate of subunits is less than $1$. \par
     Next, we theoretically derive the functional forms of the distribution of growth rates normalised by the width in the limit of $N \to \infty$. We consider three cases according to the value of $\alpha$. \textcolor{black}{The details of the derivation are given in Appendix D.}\par
   \paragraph{I: The case of $0< \alpha < 1 $:} 
The width of the growth rate does not shrink to $0$ but it converges to a finite value even in the limit of $N \to \infty,$ as known from Eq. (\ref{Eq:c8}). This behaviour can be understood by using a general property of power-law distributions with exponent less than $1$. In such a case, the mean value $\langle x_j(t)\rangle $ diverges, implying that some samples in $\left\{x_j(t)\right\}$ take extraordinarily large values compared with others. Consequently, both the denominator and numerator of Eq. (\ref{Eq:c5b}) can be approximated by only finite numbers of extraordinarily large contributors; therefore, the value of Eq. (\ref{Eq:c8}) is finite even in the limit of large $N$. The distribution of the growth rate exhibits the same property; namely, even in the limit of $N \to \infty$, the limit distribution is determined only by a small number of large contributors; therefore, we cannot expect a universal functional form in this case. \par
  \paragraph{II. The case of  $1 \leq \alpha <2$:}  
  As indicated in Fig. 2, the width of the distribution shrinks to $0$ in the limit of $N\to\infty$ and we can expect the existence of a universal limit distribution independent of the initial condition. In this case, the average, $\langle x_j(t)\rangle $, takes a finite value in the steady state, and the denominator in Eq. (\ref{Eq:c5b}) can be roughly estimated as $N\langle x_j(t)\rangle $ for very large values of $N$. However, the numerator in Eq. (\ref{Eq:c5b}) is given by the sum of $(g_j(t)-G)x_j(t)$, in which $g_j(t)-G$ gives a coefficient taking either positive or negative sign randomly with respect to $j$, and $x_j(t)$ follows a power law with  exponent $\alpha$. Namely, the numerator becomes a summation of $N$ independent identically distributed random variables that have both positive and negative power-law tails with  exponent $\alpha$. Because the generalised central limit theorem can be applied to such a sum of random variables, the limit distribution of the growth rate $\tilde{G}$, which is normalised by the width of the distribution around the mean value, is expected to converge to a stable distribution, which has the form of an inverse Fourier transform \cite{R:28} (See also Appendix C for a brief summary of the central limit theorem and  stable distributions):
\begin{equation}
p(\tilde{G};\alpha,\beta)  \equiv \frac{1}{2\pi} \int^{\infty}_{-\infty}\exp{\left\{-i \rho \tilde{G}-|\rho|^{\alpha}(1-i\beta\psi(\rho,\alpha)) \right\}} d\rho, \label{Eq:c9} 
\end{equation}
where the asymmetry parameter $\beta$, which takes a value in the interval $[-1, 1]$, and the function $\psi(\rho,\alpha)$ are given as \textcolor{black}{
\begin{equation}
            \beta = \frac{\langle (g_j-G)|g_j-G|^{\alpha-1}\rangle }{\langle |g_j-G|^{\alpha}\rangle }
, \quad \phi(\rho, \alpha) \equiv \frac{\rho}{|\rho|} \tan\frac{\pi \alpha}{2}. \label{Eq:c10}
\end{equation}
}
It is well known that the limit probability density, Eq. (\ref{Eq:c9}), has a power-law tail with  exponent $\alpha$ just like the distribution of $\left\{x_j(t)\right\}$. \par
     In Fig. \ref{fig:f3}(a), we confirm the validity of this theoretical result through a numerical simulation for the case of $\alpha=1.5$ and $\beta=0$. The normalised growth rates for the system consisting of $N$ subunits, $\tilde{G}$, are calculated by subtracting the mean value and are normalised by the width of the distribution.
   As seen from this figure the distribution of normalised growth rates changes its functional form for different values of $N$. The PDFs are  clearly converging to the theoretical function, $p(\tilde{G},1.5,0.0)$. 
   \textcolor{black}{Figure \ref{fig:f3}(b) indicates the results for the case of $\alpha=1.5$ and $\beta=0.85$. From this figure, we can also confirm convergence to the asymmetric stable distribution $p(\tilde{G},1.5,0.85)$. Such a skew distribution does not appear in the case in which $g_i(t)$ observes a normal distribution with zero mean, as discussed in Ref. \cite{wyart2003statistical}. }
     \paragraph{III. The case of $2 \leq \alpha$:}
A similar estimation for the denominator of Eq. (\ref{Eq:c5b}) is valid and the ordinary central limit theorem can be applied to the numerator of Eq. (\ref{Eq:c5b}) because the variances for $\{x_j(t)\}$ are finite.  Equation (\ref{Eq:c9}) is also valid in this case; however, the parameters are limited to $\alpha=2$ and $\beta=0$; namely, the limit distribution of the normalised growth rate is always $p(\tilde{G};2,0)$, which is the well-known Gaussian distribution with no long tail. \par 
It is interesting to consider the special situation in which the growth rates $\{g_j(t)\}$ are distributed symmetrically around $1$. 
Then, we can derive $\alpha=1$ from the basic relation $\langle g_j(t)\rangle =1$, and $\beta$ in Eq. (\ref{Eq:c10}) is $0$ by symmetry. 
In such a case, we can expect that the limit distribution of the normalised growth rate converges to  $p(\tilde{G};1,0)=1/\left\{2\pi(1+\tilde{G}^2) \right \}$ from Eq. (\ref{Eq:c9}). 
 \par
     Results for all these cases are summarised in the third column of Table 1. The limit distribution of the growth rate is determined by the value of the moment function for the growth rates of subunits. The important point is that the ordinary central limit theorem can be applied only in the limited cases of relatively small growth rates, with the mean value of the growth rate being less than $1$ and the second moment of the growth rate being less or equal to $1$. Real-world  systems are expected to be nearly in the statistically steady state and the mean values of the growth rates of subunits may take a value around $1$. Then, as one can see from Table 1, the limit distributions of the growth rates belong to either power laws or nonuniversal functional forms. \par
\section{ An application to business firm activities}
\begin{figure}[t]
 \centering
    \includegraphics[width=0.75\textwidth]{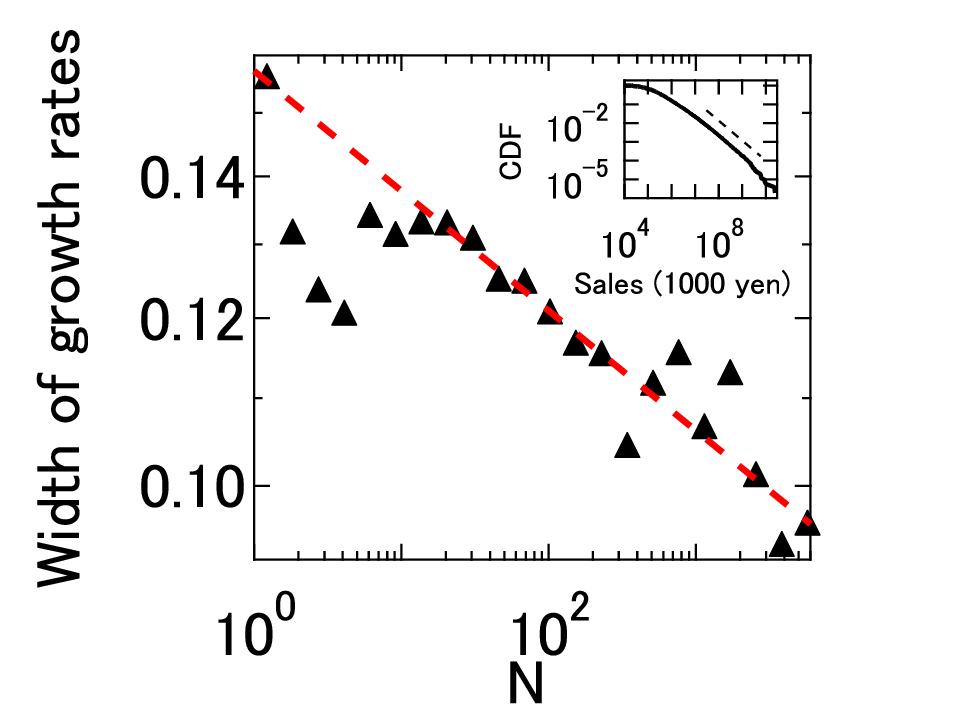}
\caption{Width of growth rate fluctuation of Japanese business firms as a function of number of workers, $N$, on a log--log plot. 
The dashed line shows a theoretical curve given by the inverse power law, \textcolor{black}{
$N^{-0.057}$, which is derived from Eq. (\ref{Eq:c8}) with $\alpha=1.06$.} The inset shows the cumulative distribution function of annual sales of about $1$ million Japanese firms in 2005 with the dotted line showing a power-law distribution with the same exponent $\alpha=1.06$. }
\label{fig:f4}       
\end{figure}
\begin{figure}[t]
   \centering
 
   \includegraphics[width=0.75\textwidth]{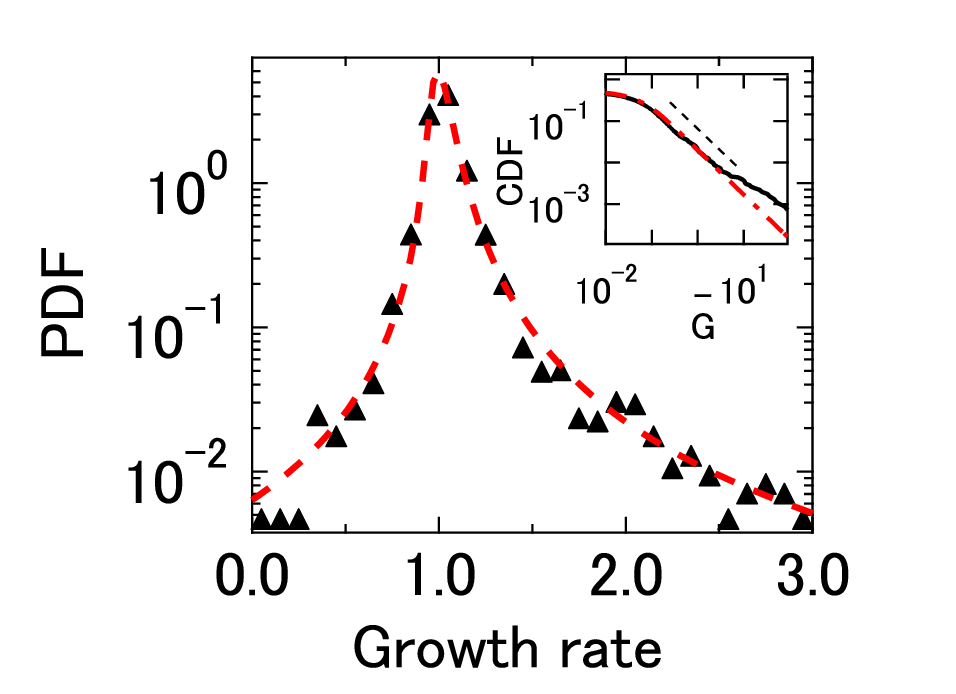}
\caption{Semi-logarithmic plot of the growth rate PDF of large business firms with more than $300$ workers (black triangles). 
The red dotted line shows the theoretical function of an asymmetric stable distribution given by $p(\tilde{G};1.06,0.50)$ in Eq. (\ref{Eq:c9}), which is translated and rescaled to fit the peak value and the width. The inset  shows the cumulative distribution function of the normalised positive growth rates (black line) on a log--log plot to confirm the fit with a theoretical stable distribution (red dashed line) and with a power law with a tail exponent $\alpha=1.06$ (black dotted line).}
\label{fig:f5}       
\end{figure}
   Now, we apply these theoretical results to data analysis. Among the various types of dynamical complex systems in the real world, business firms have been attracting the attention of scientists because there are precise observational data in the form of financial reports \cite{R:3,R:4,R:5,R:17,R:12,R:9,R:18,R:16}. To check the validity of our theory, we analyse comprehensive business firm data from Japan provided by the Research Institute for Economy, Trade and Industry (RIETI). The data consist of financial reports of 961,318 business firms, which practically cover all active firms in Japan \textcolor{black}{in 2004 and 2005}. It has already been confirmed that the basic quantities of these business firms, such as annual sales, incomes, and number of employees, all follow power-law distributions \cite{R:34}. \par
     There are several quantities that characterise the size of a business firm; these include  assets, number of employees, sales, and income. Among these quantities, we focus on sales because this quantity reflects the present scale of activity of a firm most directly. Further, we simply assume that the entire activity of a business firm is given by the sum of the activities of individual employees; namely, we regard $X(t;N)$ in Eq. (\ref{Eq:c4}) as the annual sales of a business firm with $N$ employees in the $t$th year. 
\textcolor{black}{Neglecting the additive term in Eq. (\ref{Eq:c4}) as well as the change of the number of employees in a year, we estimate the growth rate $G(t;N)$ by the ratio of the $(t+1$)th year's annual sales over that of the $t$th year's for a business firm with   $N$ employees. 
In applying  our mathematical model to real data, we need to specify the minimal independently acting subunits for actual firms. However, this is tends to be quite problematic because real firms may consist of various divisions of different sizes. To make a rough estimation, here, we simply assume that $N$ is given by the number of employees.
  It has already been confirmed from the data that the autocorrelation of the growth rate averaged over all business firms is very close to $0$, implying that the growth of a business firm can be roughly viewed as an independent random growth process approximated by Eq. (\ref{Eq:c4}) with a negligibly small external force term.} \par
     Categorising the business firms by the number of employees, $N$, we can measure the width of growth rate distribution for each category. Figure \ref{fig:f4} shows the $N$ dependence of the width of the growth rate distribution on a log--log scale. It is confirmed that Gibrat's assumption of size independence does not hold in this case, and the width of the growth rate decays clearly for large $N$.
Here, the theoretical line is given by a power law $N^{-0.046}$, \textcolor{black}{which is derived from Eq. (\ref{Eq:c8}) in the case of $\alpha=1.06$.} 
      In the inset figure the cumulative distribution of annual sales of all firms, corresponding to a superposition of the distribution of $X(t;N)$ for all $N$, is plotted on a log--log scale. We can confirm that the tail of the distribution is approximated by a power law with  exponent $\alpha=1.06$ as expected. \par   
       \par
     A theoretical limit distribution of the growth rate, $p(\tilde{G};1.06,0.50)$ in Eq. (\ref{Eq:c9}), is plotted together with that for real data estimated for $N> 300$ in Fig. \ref{fig:f5}. The inset figure shows the cumulative distribution of the normalised growth rate for positive $ \tilde{G}$ on a log--log scale to confirm the functional form of the fat tail. The growth rate distribution is asymmetric in this case and the entire functional form is well approximated by the theoretical curve of the stable distribution. This may be the first real-world example of application of an asymmetric stable distribution with a fractional  characteristic exponent value fitted in the whole range since the birth of the mathematical theory in the 1930s. \par
     \textcolor{black}{The exponent takes $\alpha_s=1.07$ [$1.063,1.068$] for the cumulative distribution function of sales, $\alpha_g=1.06$ [$1.055,1.065$]  for the cumulative distribution function of the growth rate, and $\alpha_w=0.0053=1-1/1.06$ [$1-1/1.056,1-1/1.069$] for the power law of the width of growth rates, where $[.,.]$ is the 95\% confidence interval and the domains of the power laws are assumed as $s \geq 4.22 \times 10^{10}$ (yen) for sales, $g \geq 1.11$ for the growth rate, and $\text{employees} \geq 10$ for the width.
Here, to estimate the exponents, we employed a robust linear regression after taking the logarithmic  transformation for power-law regions \cite{MaronnaMartinYohai200605}. The reason why we apply this method is to reduce the effect of outliers, which cannot be clearly distinguished from the data.  
However, it must be noted that this estimation depends on the choice of the domain of the power law, such that, for example, $\alpha_s=1.09$ [$1.086,1.089$] for $s>1.01 \times10^{11}$ (yen), $\alpha_g=1.07$ [$1.066,1.078$] for $g>1.31$, and $\alpha_w=1-1/1.06$ $[1-1/1.037,1-1/1.079]$ for $\text{employees} \geq 100$. Because of the limitation of data accuracy as well as the ambiguity of the correspondence between real business firm activity and that of the simple mathematical model, we cannot form any definite conclusions. It may be fair to conclude that the conjecture that a business firm's growth rate distribution is approximated by an asymmetric stable distribution does not contradict the data.
      }
     
     \par
\section{Discussion}
   In this paper, we introduced a theoretically solvable model of the sum of randomly growing independent subunits. As summarised in Table 1 we found generalised central limit theorems applicable for a composite of randomly growing subunits, in which we can find various types in both \textcolor{black}{width}--size relations and the limit distributions of growth rates. As an example of a real-world application, we analysed a huge database of business firm growth rates from Japan and confirmed consistency with the theory. \par
  \textcolor{black}{
     The crucial study regarding the sum of randomly growing independent subunits was investigated by Wyart and Bouchaud, and our results regarding the size dependence on the width of the PDF of growth rates are consistent therewith \cite{wyart2003statistical}. 
  Our research  clarifies the dependence of the distribution of the growth rate of the system, $G(t;N)$, on the functional form of the growth rate of the subunits. In our analysis, we consider an arbitrary functional form for the growth rate of subunits, whereas, in the case of Wyart and Bouchaud's study, only the normal distribution with zero mean is assumed for the growth rate of subunits. In particular, we clarify the categorisation of the limit behaviours with a clear condition on the growth rate distribution of subunits, as summarised in Table 1. This is a very general result that is expected to have wide application  to  randomly growing phenomena in various fields involving power laws.  
   In addition, by using our model one can derive the expression of the system growth rate $G(t;N)$ from the subunit growth rate $g_i$ more simply than by using previous models without assuming power-law distributions.
  }
   \par
     The study of applications  of this theory for growth rate distribution of complex systems is highly encouraged. It is expected that the shrinking of the growth rate width and the functional form of limit distributions can be directly compared with real data of various phenomena to check the universality of this aggregated system of randomly growing subunits. For Fig. \ref{fig:f1}, the normalised growth rate distribution looks quite similar to a double-exponential (Laplace) distribution in the case of intermediate numbers of subunits. This type of finite-size effect should be treated carefully in real-world data analysis. There is the possibility that the varieties of empirically known properties of the growth rate of complex systems introduced in the beginning of this paper can be understood by using our approach as a framework. 
  \par
     Real-world systems may not be in a steady state, so it is important in future work that transient behaviours of this independent subunit system be investigated before  the statistically steady state is reached. Extension of this novel renormalisation view of growth rates to cases of interacting subunits may also be an attractive new research topic. It is expected that a variation of the generalised central limit theorem for  growth rates might be found for the wider category of growing complex systems such as the case of nonstandard statistical physics for long-range interaction systems \cite{R:35}. \par 


%
\begin{acknowledgements}
  The authors thank RIETI for providing the business firm data. This work was partly supported by Research Foundations of the Japan Society for the Promotion of Science, Project No. 22656025 (MT), and Grant-in-Aid for JSPS Fellows No. 219685 (HW). \par
\end{acknowledgements}

\appendix

\renewcommand{\theequation}{A.\arabic{equation}}
\renewcommand{\thefigure}{A.\arabic{figure}}
\setcounter{equation}{0}
\setcounter{figure}{0}
\section*{Appendix A: Random multiplicative processes}
Because random multiplicative processes are not widely known, here we introduce a simple exactly solvable case of a random multiplicative process and show intuitively how the process realises a power-law distribution in the statistically steady state. Further, a continuum-limit version of this multiplicative process is discussed. \par
     We consider a positive random variable, $x(t)$, that follows the following stochastic equation: 
\begin{equation}
         x(t+\Delta t)=g(t)x(t)+1, \label{Eq:aa1}
\end{equation}
where $g(t)$ is a stochastic noise term that takes either a positive constant, $g$, or $0$ with probability $1/2$, respectively. Starting with the initial condition, $x(0)=1$, the time evolution is given as 
\begin{eqnarray}
x(\Delta t)= \left\{ \begin{array}{ll}
g+1 & (\text{prob}. \quad 1/2), \\
1 & (\text{prob}. \quad 1/2), \\
\end{array} \right.
\;x(2 \Delta t)=
\left\{ \begin{array}{ll}
g^2+g+1 & (\text{prob}. \quad 1/4), \\
g+1 & (\text{prob}. \quad 1/4), \\
1 & (\text{prob}. \quad 1/2), \\
\end{array} \right.
\quad  \cdots .\quad \label{Eq:aa2}
\end{eqnarray}
The general solution at time step $\tau$ is obtained as

\begin{equation}
x( \tau \Delta t)=
\left\{ \begin{array}{ll}
(g^\tau-1)/(g-1) & (\text{prob}. \quad 1/2^\tau), \\
(g^{\tau-1}-1)/(g-1) & (\text{prob}. \quad 1/2^\tau),\\
\vdots &  \\
(g^{k}-1)/(g-1) & (\text{prob}. \quad 1/2^k), \\
\vdots &  \\
1 & (\text{prob}. \quad 1/2), \\
\end{array} \right. 
\label{Eq:aa3}
\end{equation}                       
where $k$ is an integer from $1$ to $\tau$. From this solution, we can calculate the cumulative distribution of $x(t)$ in the limit of $t \to \infty$ as
\begin{equation}
P(\geq x)=2\{1+(g-1)x \}^{-\frac{\log(2)}{\log(g)}}, \label{Eq:aa4}
\end{equation}
where $P( \geq x)$ denotes the probability that $x(\infty)$ takes a value larger than or equal to $x$. In the case in which $g>1$ we have the asymptotic power-law distribution for very large value of $x$,
\begin{equation}
P(>x) \propto x^{-\alpha} ,            \label{Eq:aa5}  
 \end{equation}
where the exponent $\alpha=\log(2)/\log(g)$ fulfils Eq. (\ref{Eq:c3}) in the range $0<\alpha<\infty$, namely,
\begin{equation}
\langle g(t)^\alpha\rangle =\frac{1}{2}\cdot g^{\alpha}+\frac{1}{2}\cdot 0=1.
\label{Eq:aa6}
\end{equation}
With this special example, we confirm the validity of Eqs. (\ref{Eq:c1})--(\ref{Eq:c3}). Note that, in this example, the stationary condition, $\langle \log(g(t))\rangle <0$,  is automatically satisfied because the condition that $g(t)=0$  with probability $1/2$ gives the value $\langle \log(g(t))\rangle =-\infty$. \par 
      The key point of realising the power law in this multiplicative random process is understood intuitively by neglecting the additive term. The probability of repeating $g(t)=g$ for $k$ time steps is given as $p(k) \equiv 1/2^k= e^{-k\log(2)}$ and the corresponding value of $x(t)$ is approximated as $x(t) \approx g^k=e^{k\log(g)}$; then by deleting $k$ from these relations we have Eq. (\ref{Eq:aa5}). Namely, successive exponential growth with an exponential distribution of duration time gives the power-law distribution. \par
     This type of power-law derivation   can be generalised in the following way. Let us consider the following general form of a random multiplicative process:
\begin{equation}
x(t+\Delta t)=g(t)x(t)+f(t),  \label{Eq:aa7} 
\end{equation}
where $g(t)$ and $f(t)$ are independent random variables taking positive values, and we assume the situation in which $ \log g(t)$ fluctuates around $0$ \cite{R:40}.
By taking the logarithm of both sides and introducing variables $y(t) \equiv \log{(x(t))}$ and $r(t)=\log g(t)$, Eq. (\ref{Eq:aa7}) can be transformed as
\begin{equation}
  y(t+\Delta t)=\log{\{g(t)x(t)+f(t)\}}=y(t)+r(t)+\frac{f(t)}{g(t)x(t)}+\cdots
 .\label{Eq:aa8}
\end{equation}
By neglecting the terms including $f(t)$ as higher order terms, the time evolution of the probability density of $y(t)$, $p(y,t)$, is approximated by a Fokker-Plank equation,
\begin{equation}
  p(y,t+\Delta t) \approx \int^{\infty}_{-\infty} \omega(r) p(y-r,t)dr
=p(y,t)-\langle r\rangle \frac{\partial p(r,t)}{\partial y}+\frac{\langle r^2\rangle }{2}\frac{\partial^2 p(r,t)}{\partial^2 y}+\cdots, 
\label{Eq:aa9}  
 \end{equation}
where $\omega(r)$ denotes the probability density of $r$. Assuming the existence of a statistically steady state, we have the following exponential distribution:
\begin{equation}
        p(y) \propto \exp^{\frac{2\langle r\rangle }{\langle r^2\rangle }y} \label{Eq:aa10}
.\end{equation}
In the situation, \textcolor{black}{$\langle r\rangle =\langle \log(g)\rangle <0$}, which is equivalent to the condition of the existence of a steady state for a random multiplicative process \cite{R:31}, Eq. (\ref{Eq:aa10}) is shown to be equivalent to the power law of Eq. (\ref{Eq:aa5}), with its exponent given as
\begin{equation}
\alpha \approx -\frac{2\langle r\rangle }{\langle r^2\rangle } \label{Eq:aa11}
.\end{equation}
This value is derived from the exact relation, Eq. (\ref{Eq:c3}). Expanding the left-hand side of the equation, $\langle g(t)^\alpha\rangle =1$, in terms $\alpha$  and  equating the second and third terms on the right-hand side as an approximation lead to 
\begin{eqnarray}
   \langle g(t)^\alpha\rangle &=&\langle e^{\alpha \log(g(t))}\rangle   \nonumber \\
&=&1+\alpha \langle \log( g(t) )\rangle  + \alpha^2 \frac{\langle \{\log(g(t))^2\}\rangle }{2}+\cdots. \label{Eq:aa12}
 \end{eqnarray}
     The key equation for determining the value of the exponent, Eq. (\ref{Eq:c3}), can be derived roughly in the following way. Neglecting the additive term on the right-hand side of Eq. (\ref{Eq:aa7}) and taking an average over realisations after taking the $s$th power of both sides, we have the following relation:
\begin{equation}
\langle x(t+\Delta t)^s\rangle  \approx \langle g(t)^s\rangle  \langle x(t)^s\rangle .  \label{Eq:aa13}
\end{equation}
For $\langle g(t)^s\rangle  > 1 $  it is clear that $\langle x(t)^s\rangle $ diverges in the limit of $t \to \infty$. However, if $\langle g(t)^s\rangle  <1$ the value of $\langle x(t)^s\rangle $ is always finite. Therefore, we have the following relations: 
\begin{equation}
 \langle x(\infty)^s\rangle  =
\left\{\begin{array}{ll}
\infty & (s >\alpha ),  \\
\text{finite} & ( s< \alpha ),  \\
\end{array} \right.
      \label{Eq:aa14}                   
\end{equation}
where $\alpha$ satisfies Eq. (\ref{Eq:c3}). This property of Eq. (\ref{Eq:aa14}) is a typical characteristic of the power-law distribution, Eq. (\ref{Eq:aa5}). Therefore, we can find that the power-law exponent in Eq. (\ref{Eq:aa5}) is consistent with Eq. (\ref{Eq:c3}).  
\textcolor{black}{Note that Solomon et al. elaborated on a generalised version of this type of  model, $x(t+1)=g(t)f_1(x(t))+f_2(x(t))$, where both $f_1$ and $f_2$ are nonlinear functions, and they derived the relationship between the exponents and these functions \cite{solomon2002stable}.}
\par
     A rigorous mathematical derivation of this relation was done by Kesten in 1973 using a  more general form of a real-valued matrix by considering also the case in which the distribution of the additive noise follows a power law \cite{R:31}. In his proof the value of $\alpha$ is limited to the range  $0 < \alpha \leq 2$ as he applies the theory of stable distributions; however, our numerical analysis and the above intuitive theoretical analysis suggest that the value of $\alpha$ can be extended to the whole range $0< \alpha $. \par 
    It should be noted that the existence of the additive term in Eq. (\ref{Eq:c1}) or Eq.  (\ref{Eq:aa7}) is essential to realise the statistically steady state. As known from Eq.  (\ref{Eq:aa8}), the stochastic process is a random walk with a negative trend in view of $\log(x(t))$; therefore, without any additive noise term the random walker tends to shrink to $x(t)=0$, which is not the power-law steady state. Even without the additive term ($f(t) \equiv 0$) the same power-law steady state can also be realised by introducing a repulsive boundary condition, such as requiring $x(t) \geq 1$ for Eq. (\ref{Eq:aa7}) by adding a rule that $x(t+\Delta t)=1$ when $x(t)<1$. \par 
     Dividing both sides of Eq. (\ref{Eq:aa7}) by $\Delta t$ and considering the continuum limit of $\Delta t \to 0$, we have the following Langevin equation with a time-dependent viscosity: 
\begin{equation}
\frac{d}{dt}X(t)=-\mu(t)X(t)+F(t) \label{Eq:aa15}                                    ,         
\end{equation}
where 
\begin{equation}
X(t) \equiv x(t), \quad \mu(t) \equiv \lim_{\Delta t \to \infty}\frac{1-g(t)}{\Delta t}, \quad F(t) \equiv \lim_{\Delta t \to \infty}\frac{f(t)}{\Delta t}. \label{Eq:aa16}
\end{equation}
As known from this equation, the case of $g(t)>1$ corresponds to a negative value of viscosity ($\mu(t)<0$). In the case of a colloidal particle's diffusion in water such a negative viscosity cannot be realised; however, in the case of voltage fluctuation in an electric circuit, which is approximated also by a Langevin equation, we can consider a negative viscosity state by introducing an amplifier into the circuit. Namely, the value of $\mu(t)$ corresponds to resistivity in the electric circuit and in the situation in which fluctuation of voltage is amplified as a whole, and the effective resistivity takes a negative value. By introducing an electric circuit in which an amplifier works at random timing, we have a physical situation that is described by Eq. (\ref{Eq:aa15}) and power-law distributions of voltage fluctuation are confirmed experimentally \cite{R:36}.

\renewcommand{\theequation}{B.\arabic{equation}}
\renewcommand{\thefigure}{B-\arabic{figure}}
\setcounter{equation}{0}
\setcounter{figure}{0}
\section*{Appendix B: Basic properties of the moment function}
\begin{figure}
\centering
\includegraphics[width=6cm]{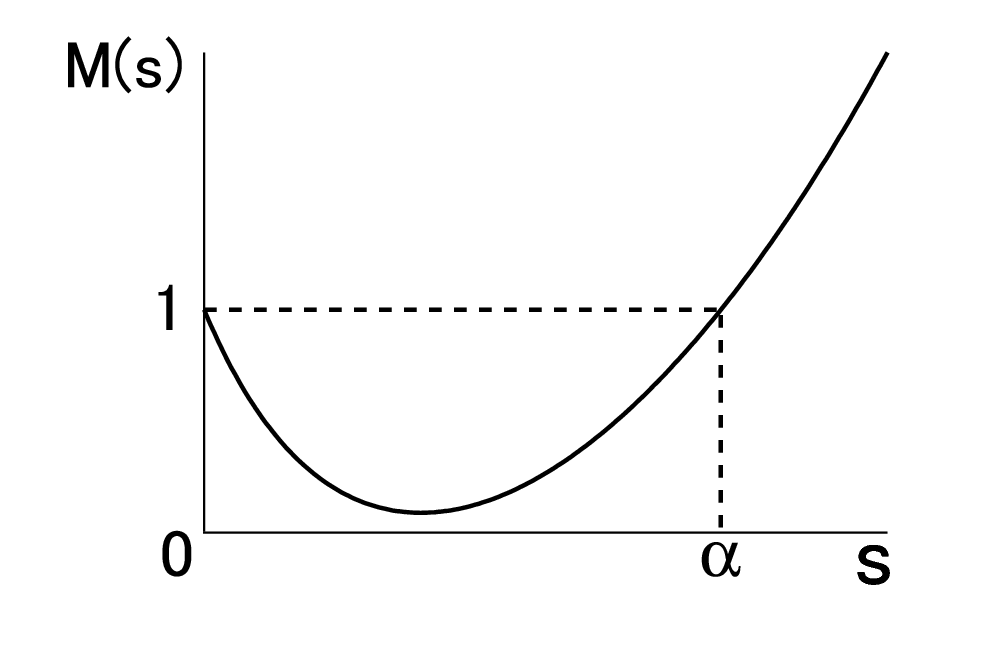}
\centering
\caption{Schematic graph of the moment function.}
\label{Fig:ab1}
\end{figure}

     Because Eq. (\ref{Eq:c3}) is  key to determining the exponent of the power law of Eq. (\ref{Eq:c2}), $\alpha$, here, we summarise the basic properties of the moment function for the growth rate of a subunit, $M(s) \equiv \langle g_j(t)^s\rangle $. 
     This is a continuous function and it is concave with respect to $s$ for any distribution of $g_j(t)$ because the second derivative of this function is always positive: $M(s)'' \equiv  \langle (\log(g_j(t))^2g_j(t)^s\rangle   >0$. Because $M(0)=1$ is an identity, if $M(\alpha)=1$ holds for a positive value of $\alpha$, then we know that $M(s)<1$ for $0<s<\alpha$ and $M(s)>1$ for $\alpha<s,$ as schematically shown in Fig. \ref{Fig:ab1}. 
    So, $M(2) < 1$ corresponds to $2< \alpha,$ as shown in the first column of Table 1.  In the situation in which Eq. (\ref{Eq:c3}) holds with $0 < \alpha <1$, then we have $M(1)=\langle g_j(t)\rangle >1$, whereas in the situation for which $1<\alpha$, we have $M(1)=\langle g_j(t)\rangle <1$. \par
    The stationary condition, $\langle \log (g(t))\rangle  < 0$, means that the slope at the origin, $M(0)'$, is  negative, so if this condition is not fulfilled $M(s)>1$ for any positive $s$, implying that the stochastic process of Eq. (\ref{Eq:c1}) is not stationary. In contrast, if the probability of occurrence of  $g_j(t) > 1$ is $0$, it is trivial that $M(s) < 1$ for any positive $s$.   
\textcolor{black}{Reference \cite{lux2002rational} provides more details and addresses applications to financial bubbles.}
\renewcommand{\theequation}{C.\arabic{equation}}
\renewcommand{\thefigure}{C.\arabic{figure}}
\setcounter{equation}{0}
\section*{Appendix C: Brief review of the generalised central limit theorem}
     The central limit theorem is one of the most powerful mathematical tools; however, it is  used too often to approximate the sum of random variables, of the form $Y(N) \equiv y_1+y_2+\cdots+y_N$, by normal distributions. In fact, there are three required conditions on random variables for their sums to obey the central limit theorem \cite{R:28}:
\begin{enumerate}  
  \item All random variables must follow an identical distribution. 
  \item The variables must be independent. 
  \item The variance of the variables must be finite.
\end{enumerate}
If one of these conditions is violated, then the central limit theorem does not apply. \par  
     Violation of the first condition has recently been attracting attention as super-statistics, that is, superposition of stochastic variables having different statistics \cite{R:37}. As an old example, a fat-tailed velocity distribution observed in randomly stirred granular particles can be explained by superposition of normal distributions having different variances owing to clustering caused by inelastic collisions \cite{R:38}. \par
     Giving a general discussion of correlated variables violating the second condition is rather difficult, as the details depend on the details of correlation. It is known that universal properties independent of the details of the system can be expected at the critical point of a phase transition at which power-law distributions and power-law scaling relations play important roles \cite{R:39}. Further, theoretical approaches based on the concept of ``nonextensive entropy'' can provide general solutions for strongly correlated systems having scale-free interactions such as charged particles \cite{R:35}. However,  when one considers activity of a business firm, for example, it seems very difficult to describe both internal and external interactions by a general mathematical formulation. \par 
     Violation of the third condition, infinite variance, was intensively studied in the 1930s by the pioneering mathematician P. Levy as ``stable distributions'' \cite{R:27}. Assuming that $\{z_1,z_2,\dots,z_N\}$  are independent identically distributed random variables, he showed that the fluctuation width of the sum $Z_N$ increases proportional to $N^{1/\alpha}$ in general, where $\alpha$ is called the characteristic exponent and lies in the range $0<\alpha \leq 2$. The limit distribution is defined for the normalised variable, $z \equiv \{Z_N-A_N\}N^{-1/\alpha}$ , where $A_N$ is a term corresponding to the mean value. In the limit of $N \to \infty$, the distribution of $z$ becomes a stable distribution that has a power-law tail $p(z)\propto z^{-\alpha-1}$ for $0<\alpha<2$. The limit distribution converges to the normal distribution when $\alpha=2$, according to the ordinary central limit theorem. This general result is called the generalised central limit theorem (GCLT). The general functional form of the stable distribution is given by Eqs. (\ref{Eq:c9}) and (\ref{Eq:c10}) \cite{R:28}. \par
\subsection*{C.I: The domain of attraction of the generalised central limit theorem}
\textcolor{black}{
Here, we review the domains of attraction of the GCLT more precisely \cite{UchaikinZolotarev199908}.
We consider the scaled sum of random variables,  
\begin{equation}
Z_N=\frac{1}{B_N}\left(\sum^{N}_{i=1}z_i-A_N\right),
\end{equation}
where $A_N$ and $B_N$ are sequences, as discussed later, and
 we assume that $z_i$ has the PDF satisfying the following conditions: 
\begin{equation}
P_{z}(x) \approx c^{+} \cdot \lambda \cdot x^{-\lambda-1} \quad (x \to \infty) \quad (j=1,2,\dots,N)
,\end{equation}
\begin{equation}
P_{z}(x) \approx c^{-} \cdot \lambda \cdot |x|^{-\lambda-1} \quad (x \to -\infty) \quad (j=1,2,\dots,N), 
\end{equation}
where $c^{+}$, $c^{-}$, and $\lambda$ are positive constants and $P_{z}(x)$ are the PDFs of $z_j$ ($j=1,2,\dots,N$).} \par
\textcolor{black}{
According to the GCLT, $Z_N$ observes a stable distribution
with the parameters 
\begin{equation}
\alpha=\begin{cases}
\lambda & (\lambda \leq 2), \\
2 & (\lambda >2),
\end{cases}
\end{equation}
and
\begin{equation}
\beta=\frac{c^{+}-c^{-}}{c^{+}+c^{-}}.
\end{equation}
Here, the characteristic function of the stable distribution is defined as
\begin{equation}
\label{stable2}
\phi(z;\alpha,\beta,\gamma,\delta)=\exp \left[i \delta z-{( \gamma |z| )}^{\alpha} \left\{ 1+i\beta\frac{z}{|z|}\omega(z,\alpha)\right \} \right]
,\end{equation}
\begin{equation}
\omega(z,\alpha)=\left\{
                \begin{array}{c}
                           \tan(\frac{\pi\alpha}{2}) \quad (\alpha \neq 1),\\
                           \frac{2}{\pi}\log(z)  \quad(\alpha =1),\\
                \end{array}
                \right.
\end{equation} 
and the coefficients $A_N$ and $B_N$ were defined as
\begin{equation}
A_N=
\begin{cases}
0 & (0 < \lambda < 1 ), \\
\beta \cdot (c^{+}+c^{-}) \cdot N \cdot \log(N) & (\lambda=1), \\
N \cdot \langle z\rangle  & (\lambda>1), \\
\end{cases}
\end{equation}
\begin{equation}
B_N=
\begin{cases}
[\pi(c^{+}+c^{-})]^{1/\alpha}\cdot[2\Gamma(\alpha)\sin(\alpha \pi/2)]^{-1/\alpha} \cdot N^{1/\alpha} & (0 < \lambda < 1 ), \\
(\pi/2) \cdot (c^{+}+c^{-}) \cdot N & (\lambda=1), \\
[\pi(c^{+}+c^{-})]^{1/\alpha}[2\Gamma(\alpha)\sin(\alpha\pi/2)]^{-1/\alpha} \cdot N^{1/\alpha} & (1<\lambda<2), \\
(c^{+}+c^{-})^{1/2}[N\log(N)]^{1/2}&(\lambda=2), \\
[(1/2)\langle (x-\langle x\rangle )^2\rangle^{1/2}] \cdot N^{1/2} &(\lambda>2). \\
\end{cases}
\end{equation}
}
\renewcommand{\theequation}{D.\arabic{equation}}
\renewcommand{\thefigure}{D-\arabic{figure}}
\setcounter{equation}{0}
\setcounter{figure}{0}
\textcolor{black}{
\section*{Appendix D: Growth rate of the system for $\alpha \geq 1$}
We calculate the growth rate $G(t;N)$ using the GCLT in the case in which the distribution converges to a stable distribution.
Here, we assume that
\begin{equation}
x_j(t+1)=g_j(t)x_j(t)+f_j(t) \quad (j=1,2,\dots,N)
\end{equation}
and for the steady state
\begin{equation}
P_{x_j}(x) \approx d_g \cdot \alpha_g \cdot x^{-\alpha_g-1} \quad (x \to \infty) \label{RMP_x} \quad (j=1,2,\dots,N)
,\end{equation}
where $d_g$ is the constant  determined by the PDF of $g_j$, $P_{g}(g)$, and  $\alpha_g$ satisfies $\langle g_j(t)^{\alpha_g}\rangle =1$. 
\subsection*{D.I: $\alpha_g > 1$} 
First, we investigate the case of $\alpha_g>1$, namely, where the 
mean of $g_j$ takes a value less than $1$.} \par
\textcolor{black}{
Considering the domain of attraction of the GCLT, as given in Appendix C, we transform $G(t;N)$ for the steady state as follows:  
\begin{eqnarray}
G_N &=&\frac{\sum^{N}_{j=1}g_j x_j}{\sum^{j=1}_{N}x_j} 
=\langle g\rangle +\frac{1}{N} \cdot \frac{B_N^{(1)}J_1}{(B_N^{(2)}/N)J_2+\langle x\rangle }, \label{App:EqEq3}
\end{eqnarray}
where $G_N$ is the growth rate for the steady state, $G_N \equiv G(t;N) \; (t \to \infty)$,  
\begin{eqnarray}
\langle g\rangle &\equiv& \langle g_j\rangle  \quad (i=1,2,\dots,N), \\
\bar{g}_j &\equiv& g_j-\langle g\rangle,  \\
\zeta_j &\equiv& \bar{g}_j \cdot x_j, \\
J_1&\equiv& \frac{\sum_{j=1}^{N}\zeta_j}{B_N^{(1)}} ,\\
J_2&\equiv& \frac{\sum_{j=1}^{N}x_j-N\langle x\rangle }{B_N^{(2)}} ,\\
B_N^{(1)} &\equiv& 
\begin{cases}
N^{1/\alpha_g} \cdot (c_1^{+}+c_1^{-})^{1/\alpha_g} \cdot F(\alpha_g)  & \text{($1 <\alpha_g < 2$),} \\
(N\log(N))^{1/2} \cdot (c_1^{+}+c_1^{-})^{1/2} & \text{($\alpha_g =2$),} \\
N^{1/2} \cdot (1/2\langle (gx-\langle gx\rangle )^2\rangle )^{1/2} & \text{($\alpha_g >2$),} 
\end{cases}   
\\
B_N^{(2)} &\equiv&
\begin{cases}
N^{1/\alpha_g} \cdot (c_2^{+}+c_2^{-})^{1/\alpha_g} \cdot F(\alpha)  & \text{($1 <\alpha_g < 2$),} \\
(N\log(N))^{1/2} \cdot (c_2^{+}+c_2^{-})^{1/2} & \text{($\alpha_g =2$),} \\
N^{1/2} \cdot (1/2\langle (x-\langle x\rangle )^2\rangle)^{1/2} & \text{($\alpha_g>2$),} 
\end{cases}
\\
F(\alpha) & \equiv& [1/\pi \cdot 2 \Gamma(\alpha)\sin(\alpha \pi/2)]^{-1/\alpha},
\end{eqnarray}
and
\begin{eqnarray}
c_1^{+}& \equiv & \lim_{y \to \infty}(1/\alpha_g) y^{\alpha_g+1} P_{\zeta_j}(y), \\
c_1^{-}& \equiv & \lim_{y \to -\infty}(1/\alpha_g) |y|^{\alpha_g+1} P_{\zeta_j}(y), \\
c_2^{+}& \equiv & \lim_{y \to \infty}(1/\alpha_g) y^{\alpha_g+1} P_{x_j}(y) ,\\
c_2^{-}& \equiv & \lim_{y \to -\infty}(1/\alpha_g) |y|^{\alpha_g+1} P_{x_j}(y). 
\end{eqnarray}
Because $B_N^{(2)}/N \to 0$ ($N \to \infty$), we can neglect the term $B_N^{2}/N \cdot J_2$ in the denominator of Eq. (\ref{App:EqEq3}) for $N\gg 1$, and we obtain the approximation 
\begin{equation}
G_N \approx \langle g\rangle +\frac{B_N^{(1)}}{N} \cdot \frac{J_1}{\langle x\rangle }.
\end{equation}
From the GCLT, $J_1$ observes a stable distribution with the parameters $\alpha=\alpha_g$, $\beta=(c_1^{+}-c_1^{-})/(c_1^{+}+c_1^{-})$, $\gamma=1,$ and $\delta=0$, where the characteristic function of the stable distribution is defined as Eq. (\ref{stable2}).
}
\par
\textcolor{black}{
Next, we specify $c_1^{+}$, $c_1^{-}$, $c_2^{+}$, and $c_2^{-}$.
By applying the formula of the transformation of random variables to $\zeta_j=b_jx_j$,
\begin{equation}
P_{(\zeta)}(z)=\int_{\zeta_{min}}^{\zeta_{max}}\frac{1}{|\bar{g}|} P_{\bar{g}}(g) \cdot P_x(z/\bar{g})dg,
\end{equation}
where $\zeta_{min}=\text{max}\{g_{min}-\langle g\rangle \}$, $\min\{z/x_{min},0\}\}$, $\zeta_{max}=\text{min}\{g_{max}-\langle g\rangle ,\text{max}\{z/x_{min},0\}\}$, the support of the PDF of $g_j$ is $[g_{min}, g_{max}]$ and the support of the PDF of $x_j$ is $[x_{min}, \infty]$.} \par
\textcolor{black}{
Taking the limit of $z$ gives 
\begin{equation}
P_{(\zeta)}(z)
=\begin{cases}
  |z|^{-\alpha_g-1} \cdot d_g \cdot \alpha_g \int_{g_{min}-\langle g\rangle }^{0}p_{\bar{g}}(g) \cdot |\bar{g}|^\alpha_g d\bar{g} & \text{$(z \to -\infty),$} \\
  |z|^{-\alpha_g-1} \cdot d_g \cdot \alpha_g \int^{g_{max}-\langle g\rangle }_{0}p_{\bar{g}}(g) \cdot |\bar{g}|^\alpha_g d\bar{g} & \text{$(z \to \infty).$}
\end{cases}
\end{equation}
}
%
\textcolor{black}{
Thus, 
\begin{eqnarray}
c_{1}^{+}&=& d_g \cdot \int_{g_{min}-\langle g\rangle }^{0}p_{\bar{g}}(g) \cdot |\bar{g}|^\alpha_g d_g \bar{g}   ,\\
c_{1}^{-}&=& d_g \cdot \int^{g_{max}-\langle g\rangle }_{0}p_{\bar{g}}(g) \cdot |\bar{g}|^\alpha_g d_g \bar{g} . 
\end{eqnarray}
In addition, from Eq. (\ref{RMP_x}), 
\begin{eqnarray}
c_{2}^{+}&=&d_g ,\\
c_{2}^{-}&=&0.
\end{eqnarray}
As a result, $G_N$ observes a stable distribution with the parameters
\begin{equation}
\alpha=
\begin{cases}
\alpha_g & \text{($1 < \alpha_g <2$),} \\
2 &  \text{($\alpha_g  \geq 2$),}
\end{cases}
\label{parameter1}
\end{equation}
\begin{equation}
\beta=\frac{c_1^{+}-c_1^{-}}{c_1^{+}+c_1^{-}}=\frac{\langle (g-\langle g\rangle )(|g-\langle g\rangle |)^{\alpha-1}\rangle}{\langle (|g-\langle g\rangle |)^\alpha\rangle }
\label{parameter2}
,\end{equation}
\begin{equation}
\gamma=\frac{B_N^{1}}{N\langle x\rangle }= 
\begin{cases}
N^{1/\alpha-1} \cdot (\langle (|g-\langle g\rangle |)^\alpha\rangle )^{1/\alpha} \cdot F(\alpha) \cdot d_g/\mu_x  & \text{($1 <\alpha_g < 2$),} \\
N^{-1/2} \cdot \log(N)^{1/2} \cdot (|g-\langle g\rangle |^\alpha)^{1/2} \cdot d_g/\mu_x & \text{($\alpha_g =2$),} \\
N^{1/2} \cdot 1/\sqrt{2} \cdot \sigma_{\zeta}/\mu_x & \text{($\alpha_g>2$),}
\end{cases}   \label{parameter3}
\end{equation}
where $\mu_x \equiv \langle x\rangle =\langle f\rangle /(1-\langle g\rangle )$ and  
$\sigma_{\zeta} \equiv \langle \{(g-\langle g\rangle )x\}^2>^{1/2}=[\langle (g-\langle g\rangle )^2\rangle\cdot \{(\langle (f-\langle f\rangle )^2\rangle+\mu_x^2 \cdot \langle (g-\langle g\rangle )^2\rangle)/(1-\langle g^2\rangle )+\mu_x^2\}]^{1/2}$.
\subsection*{D.II: $\alpha_g = 1$} 
In the same manner as was done for the case of $\alpha_g>1$, we calculate the case of $\alpha_g=1$, namely, where $\langle g_j\rangle =1$.
Considering the domain of attraction of the GCLT given in Appendix C, we transform the expression of the growth rate of the system for the steady state, $G_N$, 
\begin{eqnarray}
G_N&=&\frac{\sum^{N}_{j=1}g_j \cdot x_j}{\sum^{j=1}_{N}x_j} \\
&=&\langle g\rangle +\frac{1}{N \log(N)} \cdot \frac{B_N^{(1)} \cdot J_1+N\log(N)\cdot\beta_1 (c_1^{+}+c_1^{-})}{B_N^{(2)}/\{N \log(N)\} \cdot J_2+\beta_2 (c_2^{+}+c_2^{-}) }, \label{appappapp}
\end{eqnarray}
where 
\begin{eqnarray}
\langle g\rangle &\equiv& \langle g_j\rangle  \quad (i=1,2,\dots,N), \\
\zeta_j &\equiv& (g_j-\langle g\rangle ) \cdot x_j, \\
J_1&\equiv& \frac{\sum_{j=1}^{N}\zeta_j-N \cdot \log(N) \cdot \beta_1 \cdot (c_1^{+}+c_1^{-})}{B_N^{(1)}} ,\\
J_2&\equiv& \frac{\sum_{j=1}^{N}x_j-N \cdot \log(N) \cdot \beta_1 \cdot (c_1^{+}+c_1^{-})}{B_N^{(2)}} ,\\
B_N^{(1)} &\equiv& (\pi/2)(c_1^{+}+c_1^{-})N = (\pi/2) d_g\langle |g-\langle g\rangle |\rangle\ \cdot N
,\\
B_N^{(2)} &\equiv&  (\pi/2)(c_2^{+}+c_2^{-})N = (\pi/2) d_g \cdot N
,\\
\beta_1&=&\frac{c_1^{+}-c_1^{-}}{c_1^{+}+c_1^{-}}=\frac{\langle (g-\langle g\rangle )\rangle }{\langle |g-\langle g\rangle |\rangle }=0,\\
\beta_2&=&\frac{c_2^{+}-c_2^{-}}{c_2^{+}+c_2^{-}}=\frac{d-0}{d-0}=1 .
\end{eqnarray}
Because $B_N^{(2)}/(N\log(N)) \propto 1/\log(N) \to 0$ ($N \to \infty$), roughly, we can neglect the $B_N^{(2)}/(N \log(N)) \cdot J_2$ term in the denominator for $N\gg 1$. 
Then we obtain the following approximation:
\begin{equation}
G_N \approx \langle g\rangle +\frac{B_N^{(1)}}{N \cdot \log(N)} \cdot \frac{J_1}{d_g}.
\end{equation}
From the GCLT and the properties of a stable distribution, $G_N$ observes a stable distribution with the parameters  
\begin{eqnarray}
\alpha&=&1, \\
\beta&=&\beta_1=\frac{\langle (g-\langle g\rangle )(|g-\langle g\rangle |)^{\alpha_g-1}\rangle }{\langle (|g-\langle g\rangle |)^{\alpha_g}\rangle }=0, \\
\gamma&=&\frac{1}{\log(N)} \cdot \frac{\pi}{2} \cdot \frac{d_g \langle |g-\langle g\rangle |\rangle }{d_g}=\frac{\pi}{2} \cdot \frac{1}{\log(N)} \cdot \langle |g-1|\rangle , \\ 
\delta&=&\langle g\rangle =1.
\end{eqnarray}
This distribution is equivalent to the Cauchy distribution with the following parameters:
\begin{eqnarray}
\mu&=&1, \\
\sigma_N&=& \frac{\pi}{2} \cdot \frac{1}{\log(N)} \cdot \langle |g-1|\rangle ,
\label{parameter4}
\end{eqnarray}
where the Cauchy distribution is defined as
\begin{equation}
f_{c}(G_N;\mu,\sigma_N)=\frac{1}{\pi} \cdot \frac{\sigma_N}{(G_N-\mu)^2+\sigma_N^2}.
\end{equation}
} 
\renewcommand{\theequation}{E.\arabic{equation}}
\renewcommand{\thefigure}{E-\arabic{figure}}
\setcounter{equation}{0}
\setcounter{figure}{0}
\textcolor{black}{
\section*{Appendix E: Scaling of the variance of the model}
In Ref. \cite{wyart2003statistical},
it is reported that the scaling of the variance (or the standard deviation) 
is different from that of the PDF of the growth rates.
In an analogous way, we investigate the scaling of the variance of our model.}
\textcolor{black}{\paragraph{Theoretical approximation.}
Here, we assume that the distribution of the unit growth rates $g_i$ have the support $[0,g_{max}]$. 
It is trivial that $G_N \leq g_{max}$. 
The variance of $G_N$ is written as
\begin{eqnarray}
\langle (G_N-\langle G_N\rangle )^2\rangle  &=& \int_{0}^{g_{max}}(G_N-\langle G_N\rangle )^2 \cdot P_{G_N}(G_N) \cdot dG_N.  
\end{eqnarray}
For large $N$, the PDF of the system growth rate,  $P_{G_N}(G_N),$ is approximated by the stable distribution with the parameters
given by Eqs. (\ref{parameter1}), (\ref{parameter2}), (\ref{parameter3}), and  (\ref{parameter4}). 
On this condition, the asymptotic tail behaviour of the PDF is approximated as \cite{UchaikinZolotarev199908}
\begin{equation}
P_{G_N}(G_N) \approx \frac{1}{2}  \cdot \alpha \cdot \left(1+\beta \cdot \frac{G_N-\langle g_i\rangle }{|G_N-\langle g_i\rangle |}\right) \cdot |G_N-\langle g_i\rangle |^{-\alpha-1} \cdot N^{1-\alpha}.  
\end{equation}
Since the contribution of the central part of the PDF to the variance is much smaller than that of the tail part of the PDF (i.e. the power-law part) for large $N$, we neglect the central part.
Then we obtain
\begin{eqnarray}
\langle (G_N-\langle G_N\rangle )^2\rangle  &\approx& \int_{0}^{\langle g_i\rangle -\epsilon}(\bar{G}_N)^2 \cdot \frac{1}{2}  \cdot \alpha \cdot \left(1+\beta \cdot \frac{\bar{G}_N}{|\bar{G}_N|}\right) \cdot N^{1-\alpha} \cdot |\bar{G}_N|^{-\alpha-1} \cdot dG_N ,\\
&+& \int_{\langle g_i\rangle +\epsilon}^{g_{max}}(\bar{G}_N)^2 \cdot \frac{1}{2}  \cdot \alpha \cdot \left(1+\beta \cdot \frac{\bar{G}_N}{|\bar{G}_N|}\right) \cdot N^{1-\alpha} \cdot |\bar{G}_N|^{-\alpha-1} \cdot dG_N  \\
&\propto& N^{1-\alpha}  \label{sdsd1}
,\end{eqnarray}
where $\bar{G}_N \equiv G_N-\langle g_i\rangle $.
Therefore, the scaling of the standard deviation of the system growth rate is $ \propto N^{(1-\alpha)/2}$. 
This scaling is different from the scaling of the scale parameter of the stable distribution $N^{1-1/\alpha}$. 
}
\par
%
%
%
\textcolor{black}{
\paragraph{Numerical confirmation.}
We confirm the above-mentioned result numerically.
We calculate $G(t;N)$ for the following condition:
\begin{equation}
g_i(t)=
\begin{cases}
0   & \text{(prob 0.5),} \\
g_0 & \text{(prob 0.5}), \\
\end{cases}
\end{equation}
where $\langle g_i^\alpha\rangle =1$, namely, $g_0=2^{1/\alpha}$, and
\begin{equation}
f_i(t)=
\begin{cases}
1, & \text{ $x_i(t)< 1,$} \\
0, &  \text{ $x_i(t) \geq 1.$} \\
\end{cases}
\end{equation}
}
\par
\textcolor{black}{
For very large $N$, we transform $G_N$ as follows: 
\begin{eqnarray}
G_N&=&\frac{\sum^{N}_{i=1}g_i \cdot x_i}{\sum^{N}_{i=1}x_i} \\
&=&\frac{g_0 \cdot \sum_{i \in \{i|g_i=g_0\}}x_i}{\sum_{i \in \{i| g_i=g_0\}}x_i+\sum_{i \in \{i|g_i=0\}}x_i} \\
&=&\frac{g_0 \cdot S_a}{S_a+S_b}, 
\end{eqnarray}
where 
\begin{eqnarray}
S_a&\equiv& \sum_{i \in \{i|g_i=g_0\}}x_i, ,\\
S_b&\equiv& \sum_{i \in \{i|g_i=0\}} x_i. 
\end{eqnarray}
For $N\gg 1$, $S_a$ and $S_b$ can be approximated by a stable distribution with the parameters $\alpha=\alpha_g$, $\beta=1$, $\gamma=(N/2)^{1/\alpha_g} \cdot F(\alpha_g)$, and $\delta=N/2 \cdot \langle x_i\rangle  $ because $\sum_{i \in \{i|g_i=g_0\}}1 \sim \sum_{i \in \{i|g_i=0\}}1  \sim N/2$ and 
 $x_i$ observes the following power-law distribution:  
\begin{eqnarray}
p_{x_i}(x_i) &=&\alpha_g \cdot x_i^{-\alpha_g-1}.
\end{eqnarray}
 We apply this approximation for calculations of $G_N$ in this section.
} \par
\begin{figure}[t]
\begin{minipage}{0.45 \hsize}
\includegraphics[width=8cm]{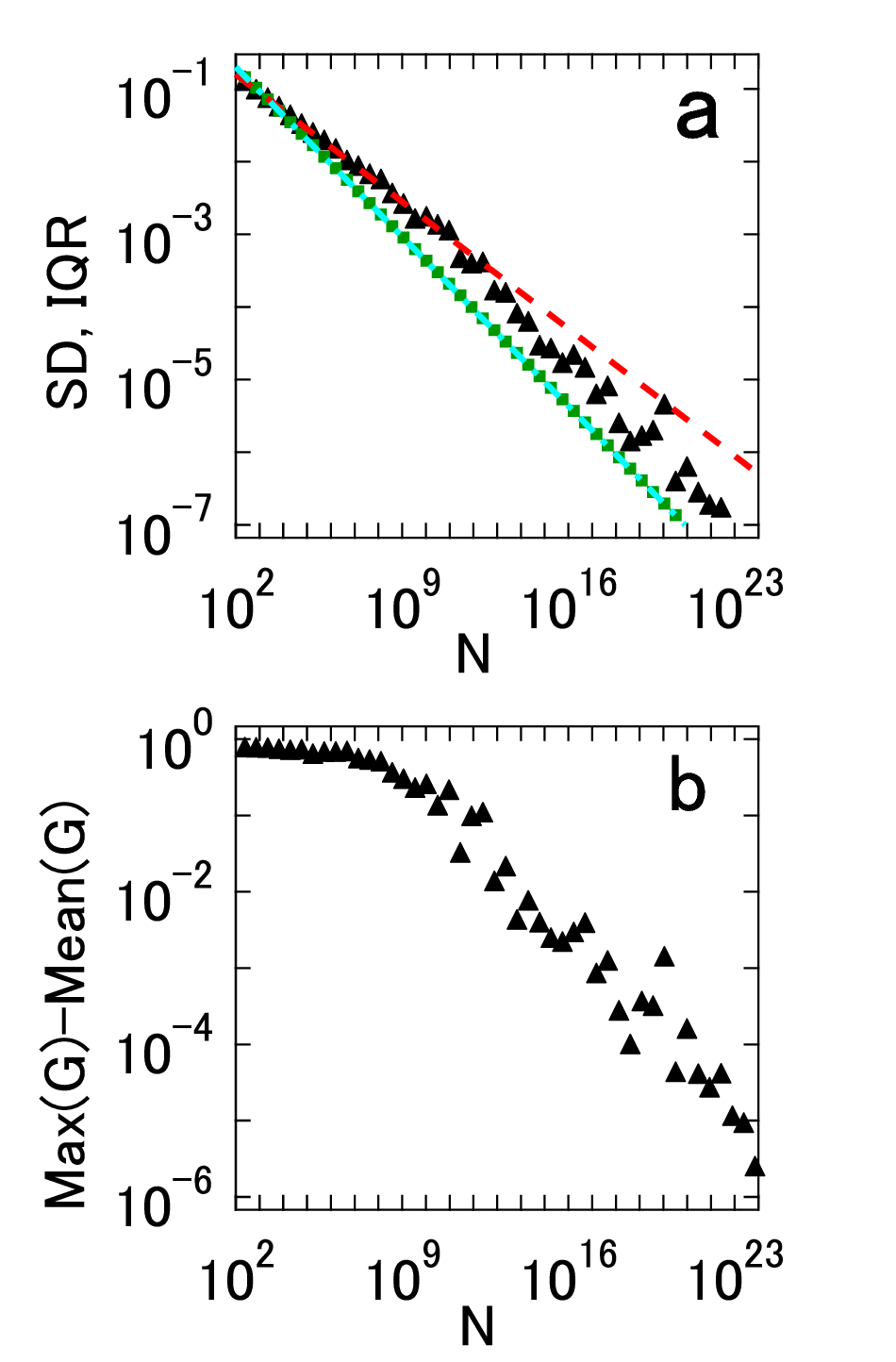}
\end{minipage}
\begin{minipage}{0.45 \hsize}
\includegraphics[width=8cm]{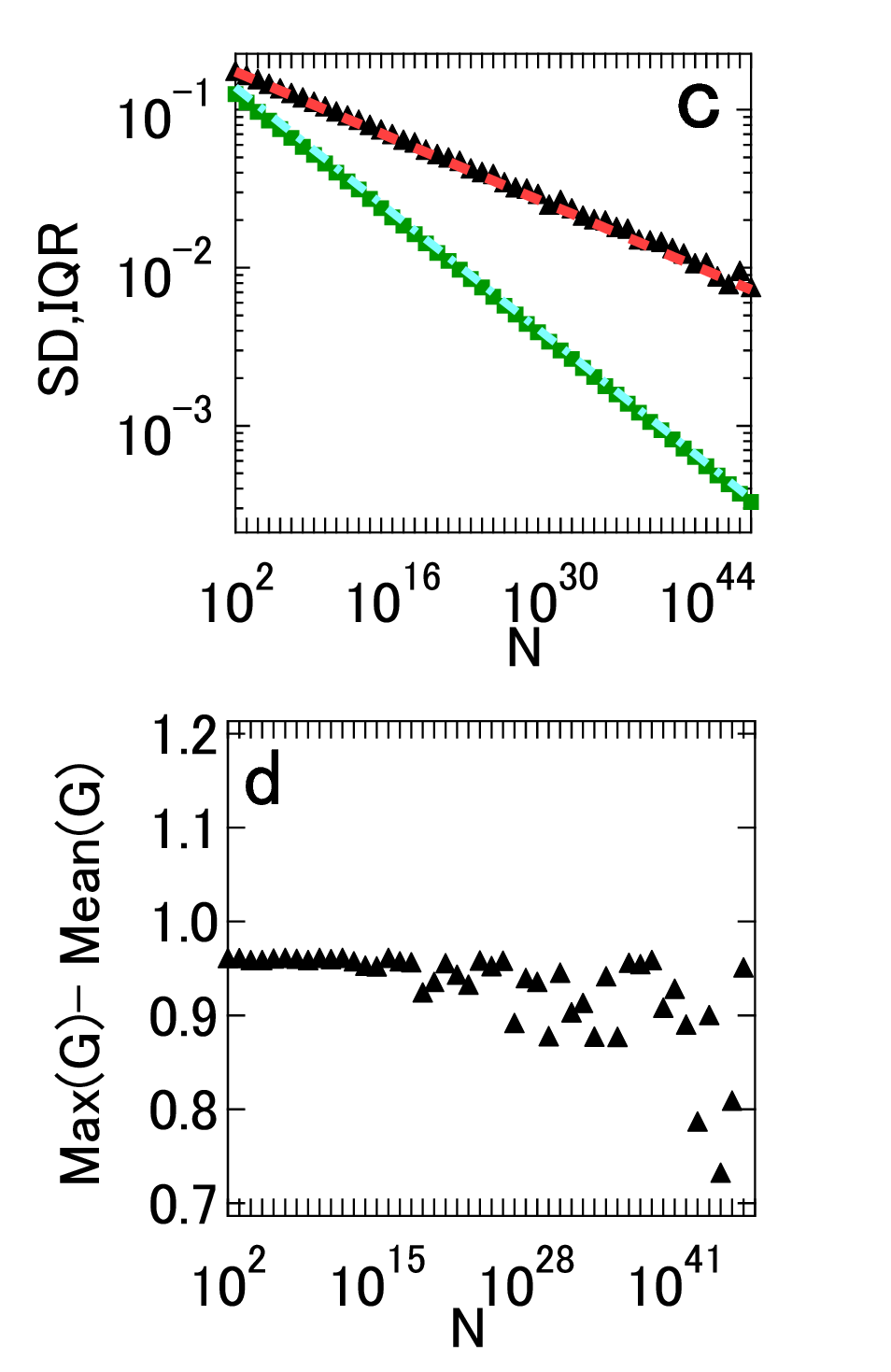}
\end{minipage}
\caption{
\textcolor{black}{Comparison of the standard deviation and the IQR for the growth rate $G_N$.
The black triangles indicate the standard deviation and the green squares indicate the IQR.  (a)  $\alpha_g=1.5;$  (b)  $\alpha_g=1.06$; red dashed lines are proportional to $ N^{(1-\alpha_g)/2}$ and blue dash dotted lines are proportional to  $ N^{1-1/\alpha_g}$. 
(c) The difference between the maximum sample of $G_N$ and the mean of $G_N$ for $\alpha_g=1.5$ and
 (d)  the corresponding figure in the case of $\alpha_g=1.06$.} }
\label{app_fig}
\end{figure}

 
 \textcolor{black}{Figures \ref{app_fig}(a) and \ref{app_fig}(b) illustrate the standard deviation and  the IQR of the system growth rate $G_N$.
From the black triangle in the figure, we confirm that the standard deviation is in accordance with the theoretical result $N^{(1-\alpha)/2}$ given by Eq. (\ref{sdsd1}) in the case of $\alpha_g=1.5$, namely, $g^{1/1.5}=1$ for $N<10^{11}$ and Eq. (\ref{sdsd1}) holds for all $N$ in the case of $\alpha_g=1.06$.} \par
\textcolor{black}{Note that in the case of $\alpha_g=1.5$ shown in Fig. \ref{app_fig}(a), for $N>10^{11}$, the standard deviation is proportional to $N^{1-1/\alpha_g}$, whose exponent is closer to that of the IQR. 
What causes  this transition of the scaling exponent from $(1-\alpha_g)/2$ to $1/\alpha_g-1$ is the finite-size effect for the sample number. 
In a finite sample, the maximum values of the asymptotic stable distribution, $g_{stable}$, are estimated at $N^{1-/\alpha_g}\cdot m^{1/\alpha_g}$ for $\alpha_g<1$ and $1/\log(N) \cdot m^{1/\alpha}$ for $\alpha_g=1$ by using the extreme value theory \cite{GumbelMathematics200407}, where we denote the sample number as $m$ (i.e. we apply $m=10^5$ in the simulation.). We can neglect the cutoff of $g_{max}$ on the condition $g_{stable}\ll g_{max}$, namely, $N^{1-/\alpha_g}\cdot m^{1/\alpha_g} \ll g_{max}$ for $1<\alpha_g<2$ or $1/\log(N) \cdot m^{1/\alpha_g} \ll g_{max}$ for $\alpha_g=1$. 
Therefore, the standard deviation observes the scaling $N^{1-1/\alpha_g}$ for $N\gg m$. 
In fact, from Figs. \ref{app_fig}(c) and \ref{app_fig}(d), we can confirm that the maximum value of the samples of $G_N$ approximately equals  $g_{max}$ for the case in which the standard deviation follows the scaling $N^{(1-\alpha_g)/2}$, namely, the case when $N<10^{11}$ for $\alpha_g=1.5$ and in the case of any $N$ for $\alpha_g=1.06$. \textcolor{black}{We should also note that 
the above-mentioned conditions that the scalings of the standard deviation are  consonant with the scalings of the IQR (i.e. $\propto N^{1-1/\alpha}$ for $\alpha \geq 1$ or $\propto 1/\log(N)$ for $\alpha>1$)  are associated with the condition that the denominator of Eq. (\ref{App:EqEq3}) can be approximated by $\langle x\rangle $ even with fluctuations considered, as $\langle x\rangle $ is sufficiently larger than the maximum value of $B_N^{(2)}/N \cdot J_2$ for a given sample number $m$.
In particular, in the case of $1<\alpha_g<2$,  the condition that the standard deviation observes $N^{1-1/\alpha_g}$  
is given by the condition $g_{stable} \approx N^{1-/\alpha_g}\cdot m^{1/\alpha_g} \ll g_{max}=\text{const}.,$ as already discussed above.
Conversely, the condition that the denominator of Eq. (\ref{App:EqEq3}) can be approximated by $\langle x\rangle $ is given by the condition $N^{1-1/\alpha_g} \cdot m^{1/\alpha}\ll   \langle x\rangle =\text{const}$. To approximate the maximum value of $J_2$, we use the approximation of $m$ samples for power-law random variables with the exponent $\alpha$ as $\{m^{1/\alpha},(m/2)^{1/\alpha},(m/3)^{1/\alpha},\dots,1\}$. 
From these calculations, we can confirm that both conditions are in accordance with the functional form of $N$.
The same discussion is applicable also for the case $\langle \alpha\rangle =1$. }} \par 
%
%

\bibliographystyle{spmpsci}      

\bibliography{growth_rev_rev_rev_7_br_rev}

\begin{thebibliography}{10}
\providecommand{\url}[1]{{#1}}
\providecommand{\urlprefix}{URL }
\expandafter\ifx\csname urlstyle\endcsname\relax
  \providecommand{\doi}[1]{DOI~\discretionary{}{}{}#1}\else
  \providecommand{\doi}{DOI~\discretionary{}{}{}\begingroup
  \urlstyle{rm}\Url}\fi

\bibitem{R:3}
Amaral, L., Buldyrev, S., Havlin, S., Leschhorn, H., Maass, P., Salinger, M.,
  Eugene~Stanley, H., Stanley, M.: Scaling behavior in economics: I. empirical
  results for company growth.
\newblock Journal de Physique I \textbf{7}, 621--633 (1997)

\bibitem{R:21}
Amaral, L., Buldyrev, S., Havlin, S., Salinger, M., Stanley, H.: Power law
  scaling for a system of interacting units with complex internal structure.
\newblock Phys. Rev.Lett. \textbf{80}, 1385--1388 (1998)

\bibitem{R:24}
Aoki, M., Yoshikawa, H.: Reconstructing Macroeconomics: A Perspective from
  Statistical Physics and Combinatorial Stochastic Processes (Japan-US Center
  UFJ Bank Monographs on International Financial Markets).
\newblock Cambridge University Press, Cambridge, Cambridge (2011)

\bibitem{R:26}
Aoyama, H., Fujiwara, Y., Ikeda, Y., Iyetomi, H., Souma, W.: Econophysics and
  Companies: Statistical Life and Death in Complex Business Networks.
\newblock Cambridge University Press, Cambridge (2011)

\bibitem{R:37}
Beck, C., Cohen, E.: Superstatistics.
\newblock Physica A: Statistical Mechanics and its Applications \textbf{322},
  267--275 (2003)

\bibitem{biham1998generic}
Biham, O., Malcai, O., Levy, M., Solomon, S.: Generic emergence of power law
  distributions and l{\'e}vy-stable intermittent fluctuations in discrete
  logistic systems.
\newblock Phys. Rev. E \textbf{58}(2), 1352 (1998)

\bibitem{R:4}
Bottazzi, G., Dosi, G., Lippi, M., Pammolli, F., Riccaboni, M.: Innovation and
  corporate growth in the evolution of the drug industry.
\newblock International Journal of Industrial Organization \textbf{19},
  1161--1187 (2001)

\bibitem{R:23}
Buldyrev, S., Growiec, J., Pammolli, F., Riccaboni, M., Stanley, H.: The growth
  of business firms: Facts and theory.
\newblock J. Eur. Econ.Assoc. \textbf{5}, 574--584 (2007)

\bibitem{R:5}
De~Fabritiis, G., Pammolli, F., Riccaboni, M.: On size and growth of business
  firms.
\newblock Physica A \textbf{324}, 38--44 (2003)

\bibitem{R:28}
Feller, W.: An Introduction to Probability Theory and Its Applications Vol. 1,
  Edition 3, volume 1 edn.
\newblock Wiley, New York, New York (1968)

\bibitem{R:6}
Fu, D., Pammolli, F., Buldyrev, S., Riccaboni, M., Matia, K., Yamasaki, K.,
  Stanley, H.: The growth of business firms: Theoretical framework and
  empirical evidence.
\newblock Proc. Nat. Acad. Sci. USA \textbf{102}, 18,801--18,806 (2005)

\bibitem{R:17}
Fujiwara, Y., Aoyama, H., Di~Guilmi, C., Souma, W., Gallegati, M.: Gibrat and
  {P}areto--{Z}ipf revisited with european firms.
\newblock Physica A \textbf{344}, 112--116 (2004)

\bibitem{R:12}
Gabaix, X.: Zipf's law for cities: an explanation.
\newblock Q. J. Econ. \textbf{114}, 739--767 (1999)

\bibitem{R:35}
Gell-Mann, M., Tsallis, C.: Nonextensive entropy: interdisciplinary
  applications.
\newblock Oxford University Press, USA, USA (2004)

\bibitem{GumbelMathematics200407}
Gumbel, E.J., Mathematics: Statistics of Extremes (Dover Books on Mathematics),
  dover edn.
\newblock Dover Publications (2004)

\bibitem{huang2002stochastic}
Huang, Z.F., Solomon, S.: Stochastic multiplicative processes for financial
  markets.
\newblock Physica A \textbf{306}, 412--422 (2002)

\bibitem{R:15}
Kalecki, M.: On the {G}ibrat {D}istribution.
\newblock Econometrica \textbf{13}, 161--170 (1945)

\bibitem{R:8}
Keitt, T., Stanley, H.: Dynamics of north american breeding bird populations.
\newblock Nature \textbf{393}, 257--260 (1998)

\bibitem{R:31}
Kesten, H.: Random difference equations and renewal theory for products of
  random matrices.
\newblock Acta Mathematica \textbf{131}, 207--248 (1973)

\bibitem{R:9}
Labra, F., Marquet, P., Bozinovic, F.: Scaling metabolic rate fluctuations.
\newblock Proc. Nat. Acad. Sci. USA \textbf{104}, 10,900--10,903 (2007)

\bibitem{R:13}
Lee, Y., Nunes~Amaral, L., Canning, D., Meyer, M., Stanley, H.: Universal
  features in the growth dynamics of complex organizations.
\newblock Phys. Rev. Lett. \textbf{81}, 3275--3278 (1998)

\bibitem{R:27}
L{\'e}vy, P., Borel, M.: Th{\'e}orie de l'addition des variables
  al{\'e}atoires, vol.~1.
\newblock Gauthier-Villars, Paris, Paris (1954)

\bibitem{lux2002rational}
Lux, T., Sornette, D.: On rational bubbles and fat tails.
\newblock Journal of Money, Credit, and Banking \textbf{34}(3), 589--610 (2002)

\bibitem{malevergne2008zipf}
Malevergne, Y., Saichev, A., Sornette, D.: Zipf's law for firms: Relevance of
  birth and death processes.
\newblock SSRN 1083962  (2008)

\bibitem{malevergne2013zipf}
Malevergne, Y., Saichev, A., Sornette, D.: Zipf's law and maximum sustainable
  growth.
\newblock Journal of Economic Dynamics and Control \textbf{37}(6), 1195--1212
  (2013)

\bibitem{MaronnaMartinYohai200605}
Maronna, R.A., Martin, D.R., Yohai, V.J.: Robust Statistics: Theory and Methods
  (Wiley Series in Probability and Statistics), 1 edn.
\newblock Wiley (2006)

\bibitem{R:7}
Mendes, R., Malacarne, L., et~al.: Statistical properties of the circulation of
  magazines and newspapers.
\newblock EPL (Europhysics Letters) \textbf{72}(5), 865 (2005)

\bibitem{R:42}
Miura, W., Takayasu, H., Takayasu, M.: Effect of coagulation of nodes in an
  evolving complex network.
\newblock Phys. Rev. Lett. \textbf{108}, 168,701 (2012)

\bibitem{R:34}
Ohnishi, T., Takayasu, H., Takayasu, M.: Hubs and authorities on japanese
  inter-firm network: Characterization of nodes in very large directed
  networks.
\newblock Prog. Theor. Phys. \textbf{179}, 157--166 (2009)

\bibitem{R:18}
Okuyama, K., Takayasu, M., Takayasu, H.: Zipf's law in income distribution of
  companies.
\newblock Physica A \textbf{269}(1), 125--131 (1999)

\bibitem{R:11}
Picoli~Jr, S., Mendes, R.: Universal features in the growth dynamics of
  religious activities.
\newblock Phys. Rev. E \textbf{77}, 036,105 (2008)

\bibitem{R:10}
Plerou, V., Amaral, L., Gopikrishnan, P., Meyer, M., Stanley, H.: Similarities
  between the growth dynamics of university research and of competitive
  economic activities.
\newblock Nature \textbf{400}, 433--437 (1999)

\bibitem{R:14}
Podobnik, B., Horvatic, D., Pammolli, F., Wang, F., Stanley, H., Grosse, I.:
  Size-dependent standard deviation for growth rates: Empirical results and
  theoretical modeling.
\newblock Phys. Rev. E \textbf{77}, 056,102 (2008)

\bibitem{R:20}
Riccaboni, M., Pammolli, F., Buldyrev, S., Ponta, L., Stanley, H.: The size
  variance relationship of business firm growth rates.
\newblock Proc. Nat. Acad. Sci. USA \textbf{105}(50), 19,595--19,600 (2008)

\bibitem{R:25}
Saichev, A., Malevergne, Y., Sornette, D.: Theory of Zipf's Law and Beyond
  (Lecture Notes in Economics and Mathematical Systems).
\newblock Springer, Heidelberg, Heidelberg (2009)

\bibitem{R:36}
Sato, A., Takayasu, H., Sawada, Y.: Power law fluctuation generator based on
  analog electrical circuit.
\newblock Fractals \textbf{8}, 219--225 (2000)

\bibitem{schwarzkopf2010cause}
Schwarzkopf, Y., Axtell, R.L., Farmer, J.D.: The cause of universality in
  growth fluctuations.
\newblock arXiv:1004.5397  (2010)

\bibitem{solomon1998stochastic}
Solomon, S.: Stochastic lotka-volterra systems of competing auto-catalytic
  agents lead generically to truncated pareto power wealth distribution,
  truncated levy distribution of market returns, clustered volatility, booms
  and craches.
\newblock cond-mat/9803367  (1998)

\bibitem{solomon2002stable}
Solomon, S., Richmond, P.: Stable power laws in variable economies;
  lotka-volterra implies pareto-zipf.
\newblock Eur. Phys. J. B. \textbf{27}(2), 257--261 (2002)

\bibitem{R:33}
Sornette, D.: Critical Phenomena in Natural Sciences: Chaos, Fractals,
  Selforganization And Disorder : Concepts And Tools (Springer Series in
  Synergetics), 2 edn.
\newblock Springer, Berlin, Berlin (2006)

\bibitem{R:40}
Sornette, D., Cont, R.: Convergent multiplicative processes repelled from zero:
  power laws and truncated power laws.
\newblock Journal de Physique I \textbf{7}(3), 431--444 (1997)

\bibitem{R:39}
Stanley, H.E.: Introduction to Phase Transitions and Critical Phenomena
  (International Series of Monographs on Physics), reprint edn.
\newblock Oxford Univ Pr on Demand, Oxford (1987)

\bibitem{R:2}
Stanley, M., Amaral, L., Buldyrev, S., Havlin, S., Leschhorn, H., Maass, P.,
  Salinger, M., Stanley, H.: Scaling behaviour in the growth of companies.
\newblock Nature \textbf{379}, 804--806 (1996)

\bibitem{R:16}
Sutton, J.: Gibrat's legacy.
\newblock Journal of economic Literature \textbf{35}, 40--59 (1997)

\bibitem{R:38}
Taguchi, Y., Takayasu, H.: Power law velocity fluctuations due to inelastic
  collisions in numerically simulated vibrated bed of powder.
\newblock Europhys. Lett. \textbf{30}, 499 (2007)

\bibitem{R:29}
Takayasu, H.: Stable distribution and levy process in fractal turbulence.
\newblock Progr. Theoret. Phys \textbf{72}, 471--479 (1984)

\bibitem{R:30}
Takayasu, H.: $f^{-\beta}${P}ower {S}pectrum and {S}table {D}istribution.
\newblock J. Phys. Soc. Japan \textbf{56}, 1257--1260 (1987)

\bibitem{R:19}
Takayasu, H., Okuyama, K.: Country dependence on company size distributions and
  a numerical model based on competition and cooperation.
\newblock Fractals \textbf{6}, 67--79 (1998)

\bibitem{R:32}
Takayasu, H., Sato, A., Takayasu, M.: Stable infinite variance fluctuations in
  randomly amplified langevin systems.
\newblock Phys. Rev. Lett. \textbf{79}, 966--969 (1997)

\bibitem{R:43}
Tamura, K., Miura, W., Takayasu, M., Takayasu, H., Kitajima, S., Goto, H.:
  Estimation of flux between interacting nodes on huge inter-firm networks.
\newblock In: International Journal of Modern Physics: Conference Series,
  vol.~16, pp. 93--104. World Scientific (2012)

\bibitem{UchaikinZolotarev199908}
Uchaikin, V.V., Zolotarev, V.M.: Chance and Stability: Stable Distributions and
  Their Applications (Modern Probability and Statistics).
\newblock V.S.P. Intl Science (1999)

\bibitem{R:1}
Vicsek, T.: Fractal Growth Phenomena: 1st Edition.
\newblock World Scientific Publishing Company, Singapore, Singapore (1989)

\bibitem{R:41}
Watanabe, H., Takayasu, H., Takayasu, M.: Biased diffusion on the japanese
  inter-firm trading network: estimation of sales from the network structure.
\newblock New J. Phys. \textbf{14}(4), {043,034} (2012)

\bibitem{wyart2003statistical}
Wyart, M., Bouchaud, J.P.: Statistical models for company growth.
\newblock Physica A \textbf{326}(1), 241--255 (2003)

\bibitem{R:22}
Yamasaki, K., Matia, K., Buldyrev, S., Fu, D., Pammolli, F., Riccaboni, M.,
  Stanley, H.: Preferential attachment and growth dynamics in complex systems.
\newblock Phys. Rev. E \textbf{74}, 035,103 (2006)

\end{thebibliography}

\end{document}